\documentclass[onecolumn]{emulateapj}

\usepackage{color}
\definecolor{blue}{rgb}{0,0.08,0.45}  

\newcommand{\be}{\begin{displaymath}}
\newcommand{\ee}{\end{displaymath}}
\newcommand{\bea}{\begin{eqnarray}}
\newcommand{\eea}{\end{eqnarray}}  

\def\vk{v_{\rm K}}

\begin{document}

\shorttitle{Magnetically Torqued Accretion Disks}
\shortauthors{Klu\'zniak and Rappaport}
\title{MAGNETICALLY TORQUED THIN ACCRETION DISKS}

\author{W. Klu\'zniak\altaffilmark{1,2} and  S. Rappaport \altaffilmark{2}}   

\altaffiltext{1}{Copernicus Astronomical Center, ul. Bartycka 18, 00-716 Warszawa, Poland; Zielona G\'ora University, ul. Lubuska 2, 65-265 Zielona G\'ora, Poland {\tt wlodek@camk.edu.pl}}
\altaffiltext{2}{Department of Physics and Kavli Institute for Astrophysics and Space Research, MIT, Cambridge, MA 02139; {\tt sar@mit.edu}}


\begin{abstract}
We compute the properties of a geometrically thin, steady accretion
disk surrounding a central rotating, magnetized star.  The
magnetosphere is assumed to entrain the disk over a wide range of
radii. The model is simplified in that we adopt two (alternate) ad
hoc, but plausible, expressions for the azimuthal component of the
magnetic field as a function of radial distance.  We find a solution
for the angular velocity profile tending to corotation close to the
central star, and smoothly matching a Keplerian curve at a radius
where the viscous stress vanishes. 
{The value of this ``transition''
radius is nearly the same for both of our adopted $B$-field models.
}
We
then solve analytically for the torques on the central star and for
the disk luminosity due to gravity {\em and magnetic torques}.
{When expressed in a dimensionless form,
the resulting quantities depend on
one parameter alone, the ratio of the
transition radius to the corotation radius.
} 
For rapid rotators, the
accretion disk may be powered mostly by spin-down of the central
star. These results are independent of the viscosity prescription in
the disk. We also solve for the disk structure for the special case
of an optically thick alpha disk. Our results are applicable to a
range of astrophysical systems including accreting neutron stars,
intermediate polar cataclysmic variables, and T Tauri systems.

\end{abstract}


\section{Introduction}
\label{sec:intro}

In a wide class of objects, from
proto-stars to neutron stars, the central accreting object is expected to
be endowed with a magnetic field of dipole strength sufficient to influence
the motion of matter in the inner parts of the accretion disk. The degree of
this influence depends on the (unknown) details of the interaction of the
dipole field with the accreting fluid.
It is not clear {\it a priori} whether the magnetosphere penetrates the disk,
or whether it is capable of transmitting significant torques if it does.
In the most commonly accepted model, the stellar magnetic field entrains 
the inner accretion disk as a result of Rayleigh-Taylor and other instabilities and exerts a torque on the disk, whose sign depends on the relative angular velocity of the disk and the star. This can result in a spin-up or spin-down torque on the star, depending on the value of the inner radius of the disk and on the accretion rate.

In this work, following the formulations of Wang (1987, 1995), Livio \& Pringle (1992),
 and others, 
we adopt a simple model of distributed magnetic torques on the disk, in which their magnitude depends on the ratio of the local orbital angular velocity to the stellar rotation rate. The underlying assumption
is that the external magnetic dipole penetrates the (thin) accretion disk for a wide range of radii (Livio \& Pringle 1992, Wang 1996). The adopted model allows us to compute the spin-up/spin-down torques on the central star, as well as the luminosity of the disk, with no reference to the viscosity law or to the actual value of the pressure and other disk variables. If, additionally, one adopts a specific prescription for the viscosity, e.g., as in an alpha disk, it is also possible to compute the detailed radial profile of the thermodynamic variables of the disk.

The inner termination radius of an accretion disk is of considerable
astrophysical interest, as it affects the flow of energy and angular momentum
in the accretion process. In the early literature, the termination radius of the disk was thought to increase smoothly across the corotation radius as the mass accretion rate drops, with matter being ejected from the system by the super-Keplerian  magnetosphere as soon as the disk is pushed out beyond the corotation radius (e.g., Davidson \& Ostriker 1973, Illarionov \& Sunayev 1975). More recently, it has been argued that the termination radius does not necessarily become larger than the corotation radius for the so-called ``fast pulsars''
{(Wang 1987), but that accretion may be stopped when ``the stellar magnetic field imparts more angular momentum to the disk plasma than is removed by internal viscous forces in an unperturbed Keplerian disk" (Wang 1995). Other authors agree that the disk terminates within the corotation radius, but argue that accretion will proceed even for ``fast pulsars," because
}
no matter is actually ejected from the disk by the rotating magnetosphere\footnote{The actual configuration of the magnetic field can only be determined by time-dependent MHD simulations (e.g., Kato et al.~2001). Recent numerical MDH calculations are exploring when   
and how the propeller mechanism operates.  Perhaps in the near
future they will provide a definitive answer as to what critical
``fastness'' parameter of the system (see \S 4, eq.~[8]) is required for
the onset of the `strong' propeller effect (e.g., Ustyugova et al.~2006).
}
(Spruit \& Taam 1993; Rappaport, Fregeau \& Spruit 2004). Our results support the latter viewpoint.

The plan of the paper is straightforward.
After briefly reviewing some of the more relevant observations, we write down the governing angular momentum
equation, we solve it, and then we discuss the implications of this solution, particularly for the torque on the central star and for the disk luminosity.  In particular, in \S 3 we present an overview of our
magnetically torqued disk solution, in \S 4.1 we discuss the adopted
torque model, and in \S 4.2 we determine the {inner} radius of
the {\em viscous} disk.  In \S 5 we compute the torques acting on the
disk as well as the concomitant luminosity.  Our
general results are summarized in graphical form in \S 6, while in \S 7 we give a detailed solution for a magnetically torqued alpha disk.  The results are
summarized and discussed in \S 8.  {An alternate $B$-field model is discussed in Appendix B, previous work is reviewed in Appendix A, and in Appendix C a model of a magnetically dominated (zero-viscosity) {\em Keplerian} disk is presented.}

\section{Estimates and observations of torques.}
\label{sec:history}

Several classes of astronomical sources involve accretion onto a rotating central star supporting a strong magnetic field. These include young
(proto)stellar objects, such as T Tauri stars, as well as accreting stellar remnants, i.e., white dwarfs (cataclysmic variables and polars) or neutron stars (accretion-powered X-ray pulsars).

It is now possible to directly measure the magnetic field strength in
the inner parts of certain proto-stellar accretion disks, such as the one around
FU Orionis (Donati et al. 2005).   Accretion torques have been studied early on for white dwarfs (e.g., Lamb \& Melia 1987). However, the most detailed information
on accretion torques comes from studies of X-ray pulsars (Bildsten et al. 1997, and references therein).
In particular, it has been found that abrupt transitions occur from spin-up to spin-down of the neutron star, with no clear change of the pulsed luminosity (which is presumed to originate in matter that is channelled to the magnetic poles), and so perhaps without a large change in the mass accretion rate. Nelson et al. (1997) take the view that in Roche-lobe overflow systems the disk may  change the sense of its rotation from prograde to retrograde, but  Li \& Wickramasinghe (1998) deem this an unlikely possibility.  
While the transitions remain unexplained, we adopt the conventional view that the disk is prograde, but the magnetosphere can transmit angular momentum of either sign between the star and the disk, depending on the location of the inner edge of the disk and the mass accretion rate.

Neither the physics of angular momentum transport by the stellar magnetosphere, nor the mechanism of penetration of the magnetic field into the accretion disk is well understood. The main uncertainty concerns the radial extent of the disk region threaded by the external magnetic field, as well as the degree to which
the magnetic field threading the disk is reduced in  magnitude by screening currents, and yet there is a remarkable agreement as to the value of the radial distance from the star at which magnetic stresses balance hydrodynamic stresses. Various authors, using very different physical assumptions, obtain rather similar values for this radius, at least for the case when it is within the corotation radius. As we show in Appendix A, because of the rapid, $1/r^6$, variation with distance  of the squared dipole field strength, and because of a universal scaling of physical quantities in a thin rotation-supported disk, detailed estimates  agree, within a factor of $\lesssim 3$, with the simplest dimensionally correct formula
\bea
r_m = \left( GM\right)^{-1/7}\dot M^{-2/7} \mu^{4/7} 
~~  \label{rm1}
\eea
(e.g., Lamb, Pethick, \& Pines 1973; Rappaport \& Joss 1976; Ghosh \& Lamb 1979; Wang 1987; Arons 1993). The modest uncertainty in the estimate of the magnetospheric radius is further reduced by a square root when one computes the ``material" torque on the star 
\bea
N_m\sim \dot M \sqrt{GMr_m} \propto \dot M^{6/7}.
\label{naive}
\eea
Observations of X-ray pulsars, indicate a steeper dependence of the spin-up torque {on} luminosity,
\bea
N\sim L^\beta
\eea
with $\beta > 6/7$ (Bildsten et al. 1997), and possibly even $\beta>1$ (Parmar et al. 1989).
This suggests that either the observed luminosity is not proportional to $\dot M$, or the torque
is given by an expression different from eq.~(\ref{naive}), or both.

In this paper, we investigate in detail the non-trivial variation with $\dot M$ of the (magnetic and material) torques exerted on the star and of the disk luminosity that follow from the adopted model of the magnetosphere--disk interaction. 
{We find that at high accretion rate $N\propto\dot M^{9/10}$ (for the alternate model considered in Appendix B,
$N\propto\dot M^{6/7}$).
}
 At low accretion rates, the torques on the central star naturally reverse sign and yield spin-down.
While the exact value of the spin-up and spin-down torques somewhat depends on the choice of the magnetic torque model, we find it to be generally true that at low accretion rates the luminosity of the disk is not proportional to the mass accretion rate. In fact, the disk luminosity is greatly enhanced by the torques on the disk whenever the star is being spun down.

%

\section{Overview of disk solution}
\label{sec:preview}

In the literature,  one can find two approaches to the problem of an accretion disk interacting with the magnetosphere of a central rotating magnetic dipole. In the first, a standard thin disk  adjusts the rotation rate of its inner edge to that of the star---and may be entrained by the external magnetic field---within a narrow boundary layer (e.g., Ichimaru 1978; Scharlemann 1978; Arons 1993). In the second, the magnetic field entrains the disk over a wider region (e.g., Ghosh \& Lamb 1979; Wang 1987; Li \& Wickramasinghe 1998; Rappaport, Fregau \& Spruit 2004), allowing a smoother variation of the torque on the central star with the mass accretion rate. Matt \& Pudritz (2005) suggest that one or the other approach may be appropriate, depending on the accretion state of the system.
For the purposes of this paper, we adopt the second approach.

The actual profile of angular velocity in the entrainment region is not known, but it is clear that the angular velocity has a maximum, at a radius $r_1$. Most authors assume that the disk is nearly Keplerian up to $r_1$, and that radius is taken to be close to the magnetospheric radius, $r_1 \simeq r_m$. The viscous torque is usually taken to vanish only at $r_1$. It is understood that this procedure is not consistent, i.e., the angular velocity cannot be both Keplerian ($\Omega \propto r^{-3/2}$ for $r\ge r_1$), and have a maximum ($d\Omega/d r =0$) at $r_1$. The inferred magnetic stress at  $r_1$ varies by a factor of four, depending on whether the former or the latter is taken to hold (e.g., Li \& Wang 1996), because the magnetic stress is taken to be proportional to $d(\Omega r^2)/d r$ (see eq.~\ref{novisc}). 

In this paper we adopt a model for the magnetic torques on the disk and then proceed to solve the azimuthal component of the equation of motion.
We assume that the viscous torque vanishes at a certain radius $r_0$, and that the disk is Keplerian at least for all $r>r_0$. This implies that $r_0>r_1$. The value of $r_0$ follows from the adopted model of the magnetic torque.
The viscous torque can vanish at $r_0$ because the magnetospheric stresses are sufficiently high to remove angular momentum (from a thin annulus at this radius) at the rate required to sustain the prescribed, constant and uniform mass accretion rate. The magnetospheric stresses on the disk can only increase inwards from $r_0$. Accordingly we take the viscous torque to vanish for all $r<r_0$ as well. No other assumptions are needed to solve for $\Omega (r)$, once the magnetic torques are specified. It is easy to find an angular velocity profile  that smoothly matches the stellar rotation rate at small radii and the Keplerian curve at $r_0$ (\S\ref{sec:match}).
 
An additional point should be mentioned.
If the stellar magnetic field entrains the disk, it exerts a torque on it, i.e., it either deposits angular momentum in the disk, or removes some angular momentum from it. This affects the amount of angular momentum transported in the disk by the viscous torques, and hence the rate of energy  dissipation and the luminosity of the disk (\S\ref{sec:viscous}).

\section{MAGNETICALLY TORQUED THIN DISKS}
\label{sec:mttd}

\subsection{Angular momentum transport}
\label{sec:angmom}

We start by defining a few fiducial radii within the accretion disk which we will utilize in this work.  The corotation radius is taken to be the radial distance at which the Keplerian angular 
frequency{, $\Omega_K$,}
is equal to the rotation frequency of the central star, $\omega_s$:
\bea
\label{rc}
r_c \equiv \left(\frac{GM}{\omega_s^2}\right)^{1/3} ~~,
\eea
where $M$ is the mass of the central star.  We also define a convenient dimensional magnetospheric radius:
\bea
r_m \equiv \left( GM\right)^{-1/7}\dot M^{-2/7} \mu^{4/7} 
~~ , 
\label{rm}
\eea
where $\dot M>0$ is the steady-state mass accretion rate,  and $\mu$ is the magnetic dipole moment of the central star.  Here and throughout this work we take eq.~(\ref{rm}) to be a formal definition of $r_m$, however, our value of $r_m$ is identical to the inner disk radius in the model of Arons (1993).  Because the ratio of these two fiducial radii appears quite often, we define a dimensionless parameter, $\xi$, which relates all four system parameters of the problem ($M, \dot M, \mu, \omega_s$):
\bea
\label{xi}
\xi \equiv \frac{r_m}{r_c} ~~.
\eea
{For a fixed system, this parameter is a measure of the mass accretion rate:
$\xi^{-7/2} \propto \dot M~$, for $M={\rm const},~\mu={\rm const},~\omega_s={\rm const}~,$
 so we may define a fiducial accretion rate $\dot M_0$ through
\bea
\label{dotM0}
\xi^{-7/2} = \dot M/\dot M_0~, ~~~{\rm for}~~M={\rm const},~\mu={\rm const},~\omega_s={\rm const}~.
\eea
}
Finally, we define $r_0$ as the radial inner boundary of the region where the accretion disk is Keplerian and the viscous torque is non-vanishing.
{The ``fastness'' parameter corresponding to this transition radius is defined as
\bea
\label{fastness}
\omega \equiv \frac{\omega_s}{\Omega_K(r_0)} = \left(\frac{r_0}{r_c}\right)^{3/2} ~~.
\eea
}

The equation governing conservation of angular momentum for a thin accretion disk subjected to distributed magnetic torques from the central star is:
\bea
-\frac{\dot M}{4 \pi Hr} \frac{d}{dr} \left(\Omega r^2 \right) = \Gamma_{\rm visc} + \Gamma_B ~ < 0 ~~,
\eea
where $r$ is the radial coordinate, $\Omega$ is the local orbital angular frequency of the disk material, $2H$ is the full disk thickness (which is a function of $r$), $\Gamma_{\rm visc}$ is the viscous torque per unit volume, and $\Gamma_B$ is the magnetic torque on the disk per unit volume.   Here, the advection of angular momentum by matter flowing inwards is driven by the viscous and magnetic torques, where a positive sign for $\Gamma_B$ indicates that angular momentum is deposited in the disk.  The vertically averaged viscous torque per unit volume can be written as:
\bea
\Gamma_{\rm visc} = \frac{1}{2Hr} \frac{d}{dr} \left( \overline{t}_{r\phi} 2H r^2 \right) = \frac{1}{2Hr} \frac{d}{dr} \left(T_{r\phi} r^2 \right)~~ ,
\eea
where $\overline{t}_{r\phi}$ is the vertically averaged-- (and $T_{r\phi}$ is the height integrated--)  $r$ - $\phi$ component of the viscous stress-energy tensor; the two are related through $T_{r\phi}=\overline{t}_{r\phi} \cdot 2H$.

We take the magnetic torque on the disk to be distributed over a range of radial distances with the magnetic torque per unit volume given by:
\bea
\Gamma_B = {r \over 4 \pi}\frac{\partial (B_z B_\phi)}{\partial z}  ~ ,
\eea
so that the vertically averaged magnetic torque is
\bea
\frac{1}{2H}\int \Gamma_B\, dz =\frac{B_z B_\phi r}{4 \pi H}  ~ ,
\eea
where, in the last term, the magnetic field is evaluated at the top ``surface'' of the disk, i.e., at
$z=H$.

Without a full, self-consistent magnetohydrodynamic solution of the disk equations, there is no simple way of determining $B
$.  In principle, the magnetic torque on the accretion disk should be computed self consistently, but for purposes of gaining insight into the global effects of magnetic torques we prefer to adopt a reasonable analytic model rather than working with a more complex problem that is not necessarily any more valid.  To obtain results that would be applicable to any
thin disk model, we simply adopt the following somewhat {\em ad hoc} but physically plausible prescription for $B_\phi$ at $z=H$ (Livio \& Pringle 1992; Wang 1995; Rappaport, Fregeau, \& Spruit 2004):
\bea
\label{bphi}
B_\phi = B_\phi^I \simeq  B_z \left( 1 - \frac{\Omega}{\omega_s} \right) ~ ,
\eea
which we take to hold for $r \ge r_c$ (see also, Wang 1987; 1995; 1996 for more sophisticated and physically motivated versions of this relation).  Here and in the following we assume, without loss of generality, that $B_z >0$.  The expression for $B_\phi$ given in eq.~(\ref{bphi}) has the property that the magnetic torque vanishes at the corotation radius, while $B_\phi$ is comparable to $B_z$ at large distances. The latter is compatible with the fact that $B_\phi$ cannot be larger than $B_z$ over significant distances, for reasons of equilibrium and stability of the field above the disk plane (see discussion in Rappaport et al. 2004).  As the field is wound up by differential rotation  between the star and the disk, the azimuthal component first increases, but the increasing energy in the azimuthal field component pushes the field configuration outward into an open configuration when the azimuthal component becomes comparable to the poloidal component. This was proven by Aly (1984, 1985) in a rather general context, and worked out in some detail for the case of a disk around a magnetic star by Lynden-Bell \& Boily (1994).  For radial distances inside the corotation radius ($r<r_c$) we adopt one of two plausible expressions for $B_\phi$:
\bea
\label{recipe}
B_\phi^I \simeq  B_z \left( 1 - \frac{\Omega}{\omega_s} \right) ~~{\rm or}~~B_\phi^{II} \simeq - B_z \left( 1 - \frac{\omega_s}{\Omega} \right),
\eea
where the first version, $B_\phi^I$, is just an analytic continuation of the expression given in eq.~(\ref{bphi}), while the second version, $B_\phi^{II}$, has the aesthetic advantage that $|B_\phi| \rightarrow |B_z|$ as $r \rightarrow 0$, as opposed to becoming progressively $\gg B_z$ { (for Keplerian $\Omega$)}.  However, $B_\phi^{II}$ has the disadvantage that the functional form for the magnetic torque must be switched at $r=r_c$.  We utilize both of these functions in this work, but give results in the text for $B_\phi^I$ only since these are algebraically less messy.  The corresponding results for $B_\phi^{II}$ are given in Appendix B. We note that a form similar to $B_\phi^{II}$ was discussed extensively by Erkut \& Alpar (2004), who also derive an equation very similar to our eq.~(\ref{TrphiII}).

{ These two prescriptions (eq.~[\ref{recipe}]) for the azimuthal field can be discussed in terms of the magnetic diffusion coefficient $\eta$. As argued by Livio \& Pringle (1992) the azimuthal field is created by vertical shearing. The poloidal field is stretched by a factor of $2\pi r/H$ in a time $2\pi/|\Omega -\omega_s|$, i.e., the azimuthal field is created at the rate $(r/H)|B_z(\Omega -\omega_s)|$. In steady state this has to be balanced by the diffusion rate $\eta d^2 B_\phi/d^2z\sim \eta  B_\phi/H^2$, resulting in $B_\phi\sim rHB_z(\omega_s -\Omega)/\eta$. Campbell (1992) derived the last relation from the induction equation, while Wang (1987) obtained it from considerations of buoyancy. Our two prescriptions for $B_\phi$ then correspond to $\eta \sim  rH\omega_s$, or  $\eta \sim rH\Omega$, respectively. Close to the corotation radius both prescriptions coincide. In the second prescription, the diffusivity is larger by a factor $r/H$ than the largest possible value of the coefficient of kinematic viscosity in subsonic turbulence, $\nu_{\rm max}\sim c_sH\sim H^2\Omega\sim \eta H/r$, where $c_s$ is the speed of sound, i.e., we do not necessarily assume that the magnetic diffusivity and the  disk viscosity are of the same origin (cf. Campbell 2000).

We note that Agapitou \& Papaloizou (2000) use a diffusivity prescription resulting in a torque that differs from our prescription by a constant factor $C$ (their eq.~[14]). Their discussion suggests that a large value of $C$ is more physical than the value $C=1$. However, the effect of smaller assumed diffusivity seems to be offset by the effects of an inflated field (see below).
{We further note that in the following, a rescaling of the diffusivity by a constant factor $1/\gamma_d$ corresponds to rescaling of $r_m^{7/2}$ by $\gamma_d$, and of $\mu$ by $\sqrt\gamma_d$. Wang (1995) showed that the torque on the central star is independent of $\gamma_d$, when expressed in units of $\dot M\sqrt{GMr_0}$ (see eq.~[\ref{dimlesstorqueIalt}], where we have neglected the contribution $- \omega_s R_s^2$ in eq.~[\ref{taus}]).
}
} 

{ With the azimuthal field given by $B_\phi^I $ in eqs.~(\ref{bphi}), (\ref{recipe})}
the complete height-averaged angular momentum equation now reads:
\bea
\label{visc1}
- \frac{\dot M}{4\pi Hr} \frac{d}{dr} \left(\Omega r^2 \right) = \frac{1}{2Hr} \frac{d}{dr} \left(T_{r\phi} r^2 \right) 
+ \frac{B_z^2 r}{4\pi H} \left( 1-\Omega/\omega_s \right).\nonumber\\
\eea
%

\begin{figure}[h]
\begin{minipage}[c]{0.47\textwidth}
\begin{center}
\includegraphics[height=3.0in]{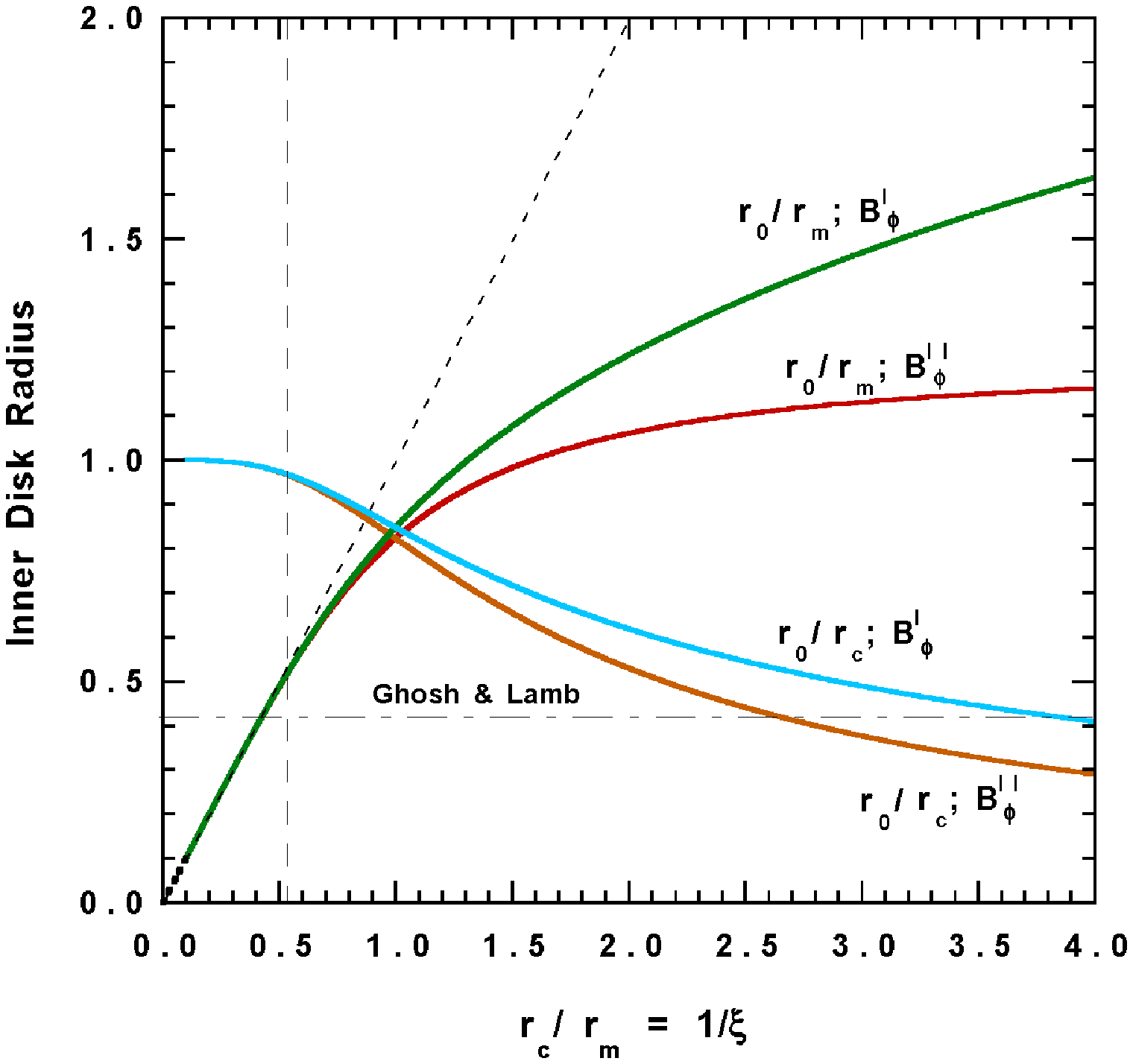}  
\caption{Ratios of  $r_0$, the inner radius of the viscous disk, to the corotation radius, $r_c$, and to the magnetic fiducial radius, $r_m$, as functions of the parameter $\xi^{-1} \equiv r_c/r_m$.  The results are shown for the two different prescriptions for $B_\phi$ (see text). 
{The vertical line indicates the value of $1/\xi$ corresponding to zero torque on the central star (eq.~[\ref{taus}]).}  
The horizontal dot-dashed curve is the Ghosh \& Lamb (1979) value for
$r_0/r_m$.\label{fig:rad}}
\end{center}
\end{minipage}
\hfill
\begin{minipage}[c]{0.47\textwidth}
\begin{center}
\null{}\vglue-1.3cm
\includegraphics[height=3.0in]{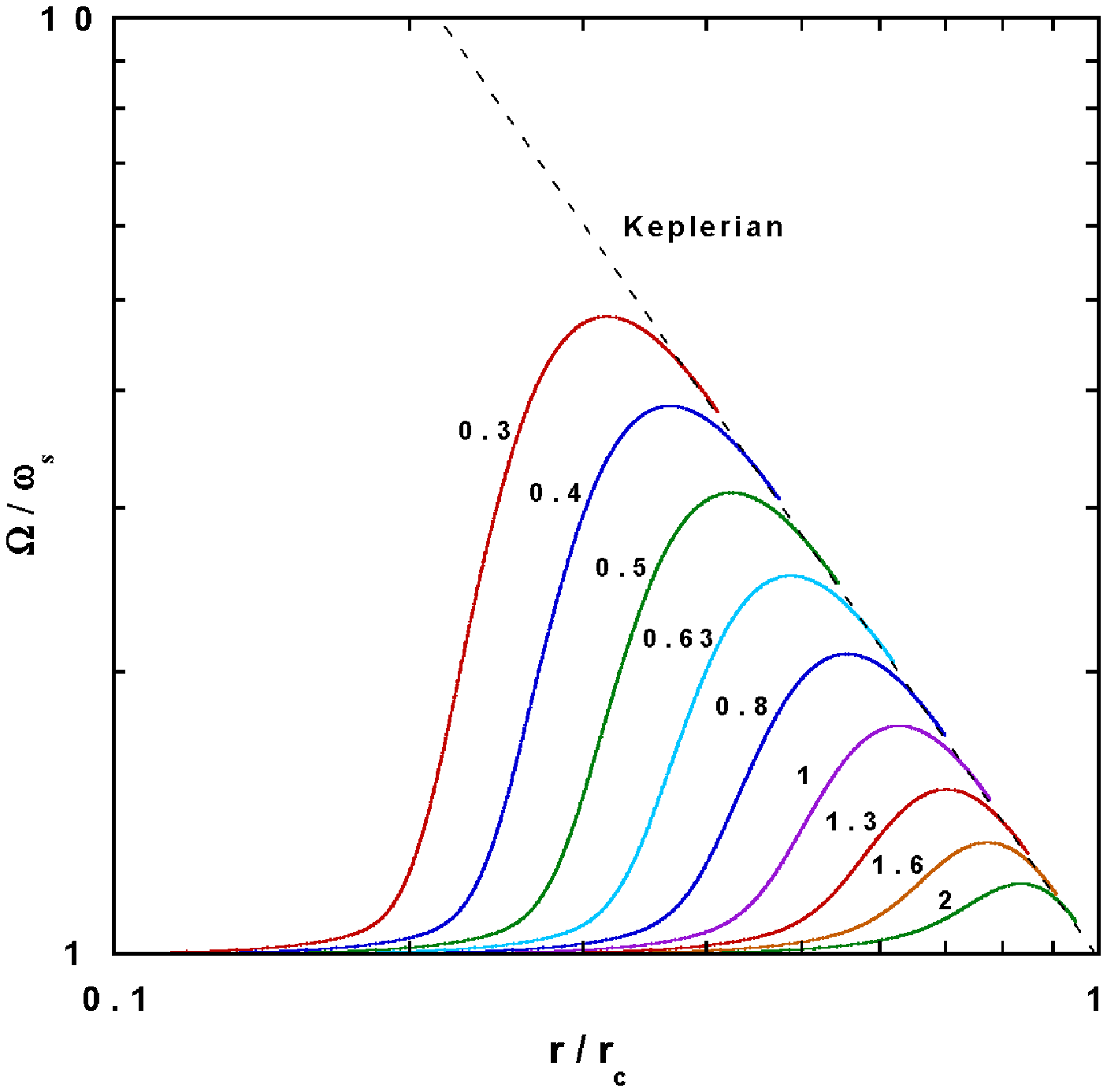} 
\caption{The matching solutions in terms of $\Omega/\omega_s$ in 
the sub-Keplerian regime as found from eq.~(\ref{novisc}).  The curves are 
labeled by the values of the parameter $\xi$.
\label{fig:matching}}
\end{center}
\end{minipage}
\vglue0.4cm
\end{figure}  

{ We consider a central dipole aligned with the stellar spin axis, and an accretion disk that is axisymmetric and perpendicular to the spin axis.
In the following we will assume that the poloidal field is given by the dipole formula $B_z=\mu/r^3$.   We expect that the results obtained in this paper will not differ qualitatively if this assumption is relaxed. For example,
the expression of Arons (1993) for the inner radius of the disk, $r_m$, which we reproduce in eq.~(\ref{Aronsr}) of Appendix A, is based on an exact solution of Aly (1980) for the field structure in the presence of a diamagnetic disk (in which, close to the inner edge of the disk the field is enhanced by a factor of $r/H$), and yet, it agrees with our result for $r_0$:
{$r_0/r_m\sim{\cal O}(1)$} for $r_m<r_c$; see eqs.~(\ref{bc}), (\ref{bcbis}), and Fig.~\ref{fig:rad}. 
Likewise, considering numerical solutions of realistic disks, Bardou \& Heyvaerts (1996) and Agapitou \& Papaloizou (2000) show that the field lines may be inflated outside the corotation radius, leading to a reduction in the value of the poloidal field relative to the dipole value that is assumed here. However, this does not seem to strongly affect the results discussed in this paper. 
 Eventually, Agapitou \& Papaloizou (2000) find values of torque which are the same as ours, within a factor of a few [see our \S \ref{sec:torques}].
}

\subsection{Matching solutions in the sub-Keplerian regime}
\label{sec:match}

Unlike most other workers, to describe the steady disk we use the same disk equation for all radii (eq.~[\ref{visc1}]). Since the model torques depend on the radial profile of the angular velocity in the disk, and the angular velocity profile depends on the torques, the azimuthal equation of motion of the accreting fluid must be solved in a self-consistent manner. Assuming no mass loss from the disk, we find a solution which asymptotically matches the rotation rate of the star at small radii, and smoothly matches a viscous Keplerian disk at a certain radius, $r_0$. The value of this radius also follows from the same eq.~(\ref{visc1}). 

{ We begin by arguing that the disk may be taken to be Keplerian where the viscous torque is non-vanishing, with the corollary that $T_{r\phi}=0$ for $r\le r_0$.
}
In a steady, thin accretion disk, the radial pressure gradients are a factor of $(H/r)^2$ smaller than gravity (Shakura and Sunyaev 1973; $H/r<< 1$ is the dimensionless disk thickness), and hence, in the absence of strong external disturbance the disk is rotation supported, i.e., nearly Keplerian everywhere. On the other hand, if the disk extends within the corotation radius, its angular velocity must reach a maximum (at a smaller value $r_1$) before it can match the lower angular velocity of the star.  At the maximum, the viscous torque must vanish. We expect the viscous torques to be zero also for all $r<r_1$; this is consistent with the fact that magneto-rotational instability (Balbus \& Hawley 1991), which is thought to be responsible for the presence of an effective viscosity in the disk, does not operate in the region of radially increasing angular velocity $d\Omega/dr>0$.
{ Hence, the viscous torque vanishes for all $r\le r_2$, for a certain $r_2\ge r_1$.
}

{ Accretion can proceed in the region $r\le r_2$ only because the magnetic torques are sufficiently high to remove angular momentum at the requisite rate. In fact, by the argument in the previous paragraph, in the sub-Keplerian region the magnetic torque must be dominant, the viscous term in eq.~(\ref{visc1}) only  becoming important when 
$|B_z^2 r \left( 1-\Omega_K/\omega_s \right)| < |-\dot M\Omega_K/2|$,
where $\Omega_K$ is the Keplerian orbital frequency. To keep the algebra clean we assume that $r_2\ge r_0$, i.e., the viscous torque vanishes already in the Keplerian region. Since eq.~(\ref{visc1}) does not admit Keplerian solutions for $T_{r\phi}\equiv 0$, we must then have $r_2 = r_0$. This is not a very restrictive assumption. If, instead of assuming $T_{r\phi}(r_0)= 0$, we were to take the viscous torque to remove one-half of the angular momentum at the matching boundary,
 $d(T_{r\phi}r^2)/dr|_{r_0}= -(\dot M /4\pi)d(\Omega_Kr^2)/dr|_{r_0}$,
the estimate of $r_0$ in eqs.~(\ref{bc}) - (\ref{r0}) would change by a factor of $2^{1/5}$ for $\xi<<1$, i.e., by less than 15\%,
and not at all in the limit $\xi {\rightarrow\infty}$. In short, we assume $T_{r\phi}\ne 0$ if and only if $r>r_0$.
}

The equation for angular momentum transport  (eq.~[\ref{visc1}]) in the region $r \le r_0$, where viscous stresses are zero, is our starting point for exploring formal flow solutions which make the transition from Keplerian rotation at $r_0$ to corotation with the central star as $r \rightarrow R_s$:
\bea
\label{novisc}
\frac{\dot M }{4\pi Hr} \frac{d}{dr} \left(\Omega r^2 \right) = - \frac{B_z^2 r}{4\pi H} \left( 1-\Omega/\omega_s \right) ~~ .
\eea
This being a first order ordinary differential equation, a single initial condition (at a fixed radius) specifies the solution. However, using the freedom of choosing the matching radius, it is possible to match both the slope and the value of the solution with a Keplerian one. Indeed, we first find the radius $r_0$ at which the solution matches  Keplerian rotation by substituting $\Omega_K r/2$ for the derivative on the left hand side, and then use
 $\Omega(r_0) = \Omega_K(r_0)\equiv \sqrt{GM/r_0^3}$ as the single initial condition to solve for $\Omega(r)$ for all $r\le r_0$.


With the described substitution, eq.~(\ref{novisc}) reduces to an algebraic equation for $r_0$:
\bea
\label{bc}
\frac{1}{2} = \left(\frac{r_m}{r_0}\right)^{7/2} \!\! \left(\sqrt{\frac{r_c^3}{r_0^3}}-1\right) = 
{\xi^{7/2} \omega^{-10/3}\,\,(1-\omega).~
}
\eea
This is essentially the same as the expression for the inner edge of the disk found by Wang (1995), and also used by Kenyon et al. (1996).  For any choice of value for the dimensionless parameter $\xi$ this equation can be solved numerically for the inner radius, $r_0$ in units of $r_c$.  The asymptotic limits for large and small values of $\xi$ are:
\bea
\label{r0hiM}
\frac{r_0}{r_c} & \simeq 2^{1/5} \xi^{7/10} ~~~~{\rm for}~~\xi \ll 1~, \\
\label{r0}
\frac{r_0}{r_c} & \simeq 1-\frac{1}{3} \xi^{-7/2} ~~~~{\rm for}~~\xi \gg 1~.
\eea
Note that $r_0\le r_c$, always. Plots of $r_0/r_c$ and $r_0/r_m$ are shown in Fig.~\ref{fig:rad}. 
{For a given system, the high accretion rate limit of  eq.~(\ref{bc}) is $r_0\propto\dot M^{-1/5}$, by eqs.~(\ref{dotM0}), (\ref{r0hiM}).
Of course, the disk cannot penetrate the surface of the star or extend far within the marginally stable orbit (ISCO) predicted by general relativity
(e.g., Klu\'zniak \& Wagoner 1995), so eqs.~(\ref{bc}-\ref{r0}) and the solutions discussed below are valid only for $r_0>>\,{\rm max}(R_s, r_{\rm ms})$.
}

{For any non-vanishing mass accretion rate, $\dot M\ne0$, eq.~(\ref{bc}) allows us to eliminate the unobservable quantity $r_m$ from all the equations, so that the ratio $r_0/r_c$ is the only remaining parameter in the expressions for various physical quantities. Specifically, the first equality in eq.~(\ref{bc}) can be rewritten as
\bea
\label{bcfast}
\left(\frac{r_m}{r_0}\right)^{7/2} \!\! = \frac{\omega}{2(1-\omega)}~.
\eea
The limit of no magnetic field corresponds to $\mu^2=0$, $r_m=0$, $\xi=0$, $\omega=0$. The limit of $\dot M \rightarrow 0$ corresponds to $\omega \rightarrow 1$, and more specifically $\dot M/(1-\omega) \rightarrow 2\dot M_0$.
}

{ The counterpart of eq.~(\ref{bc}) for our 
(\textcolor{black}{alternate}) prescription $B^{II}_\phi$ (eq.~[\ref{bcbis}] in Appendix B) has also been derived by Matthews et al. (2005). However, their interpretation of $r_0$ is different from ours. For Matthews et al., $r_0$ (their $R_t$) is the inner edge of the magnetically dominated disk, where the surface density vanishes (i.e., $\Sigma\equiv0$ for $r<R_t$), and where the radial inflow velocity is large. In our approach, $r_0$ is the {\it outer} edge of the magnetically dominated inner disk; the disk is Keplerian at $r_0$, and therefore the radial velocity must still be much smaller than \textcolor{black}{the value of} $r\Omega_K$ {at this radius}. Furthermore, there is no physical reason for the density to vanish at $r_0$.
}

The specific angular momentum of the matter flow, $\ell = \Omega r^2$, is governed by the equation valid for all $r\le r_0$ (essentially a rewritten version of eq.~[\ref{novisc}]):
\bea
\label{max}
\frac{\dot M}{r}  \frac{d\ell}{dr} = - B_z^2 r \left( 1-\frac{\ell}{\omega_s r^2} \right) ~~ ,
\eea
which  can be cast in completely dimensionless form by using rescaled variables $\mathcal{L}\equiv \ell/\sqrt{GMr_c}$, and $\mathcal{R}\equiv r/r_c$
\bea
\frac{d\mathcal{L}}{d\mathcal{R}} = - \frac{\xi^{7/2}}{\mathcal{R}^4} \left(1-\frac{\mathcal{L}}{\mathcal{R}^2} \right) ~~.
\eea
This equation admits to analytic solutions which involve incomplete gamma functions, but instead of reproducing the formulae we exhibit the solution graphically, after simply integrating this equation numerically.  The results are shown in Fig.~\ref{fig:matching}, together with the Keplerian curve.  Note that the solutions smoothly match a Keplerian $\Omega(r)$ at $r_0$, reach a maximum at a lower value of $r_1$, and finally decrease smoothly until they level off asymptotically toward $\omega_s$.

Neither the value of the maximum angular frequency, $\Omega (r_1)$, nor the value of the radius $r_1$ at which this maximum is attained, plays any role in the following considerations. This is because the viscous torque is non-zero only for radii $r>r_0>r_1$. There is no viscous dissipation, and no angular momentum can be transmitted  upstream through the fluid, for any $r<r_0$. Once the accreting fluid passes through the (imaginary) cylinder at $r_0$, all its angular momentum is transmitted to the star, at the rate $\dot M\sqrt{GMr_0}$.


\section{Torques and Luminosity}

Now that we have found a solution in which the viscous torque vanishes at
$r_0$ and the disk is Keplerian for all $r \ge r_0$ we can use eq.~(\ref{visc1}) to compute the viscous torque in the disk, and the luminosity of the disk. We are also in a position to compute the total torque on the star. These results follow from our model assumptions alone, and are
independent of the detailed structure of the thin accretion disk
(equation of state of the fluid, the viscosity prescription, disk height, etc.).

\subsection{Viscous Torque}
\label{sec:viscoustorque}
 
If we integrate both sides of eq.~(\ref{visc1}) with respect to $r$, starting at $r_0$, the largest radius where $T_{r\phi}=0$ (inner edge of the {\it viscous} accretion disk), we find:
\bea
\label{govern}
- \dot M \left(\Omega r^2-\Omega_0 r_0^2 \right) = 2 \pi T_{r\phi} r^2 + \int_{r_0}^r \frac{\mu^2}{r^4} \left 
         (1- \frac{\Omega}{\omega_s} \right) dr ,~~~\nonumber\\
\eea
where for the $z$ component of the magnetic field we have taken $\mu/r^3$ to represent an unscreened dipole field whose axis is aligned approximately along the spin axis of the central star and perpendicular to the accretion disk. 
Finally, we can carry out the remaining integral and solve for the vertically integrated $r-\phi$ component of the stress tensor at an arbitrary radial distance:
\bea
\label{visc2}
- T_{r\phi} = \frac{\dot M\sqrt{GM}}{2\pi r^2} \left(\sqrt{r} -\sqrt{r_0} \right) + \frac{\mu^2}{18 \pi r^2}    \left[\frac{3}{r_0^3}-\frac{3}{r^3}-2\sqrt{\frac{r_c^3}{r_0^9}}+2\sqrt{\frac{r_c^3}{r_{\null}^9}}\right]~~ .
\eea
Here we have made use of our result from Section~\ref{sec:match} that the disk matter is in Keplerian orbital motion everywhere for $r \ge r_0$. 

\subsection{Viscous Heating of the Disk}
\label{sec:viscous}

We now utilize our result for the vertically integrated stress to compute the local viscous heating in the disk.  The viscous heating per unit volume is given by $\dot Q = \overline{t}_{r\phi} r d\Omega/dr$, and the corresponding viscous heating per radial interval is $dL_{\rm vis}/dr = 2\pi r^2T_{r\phi} d\Omega/dr$. From this we find:
\bea
\label{lvis}
\frac{dL_{\rm vis}}{dr} = \frac{3GM\dot M}{2r^2} \left(1 -\sqrt{\frac{r_0}{r}} \right) + \frac{\mu^2\sqrt{GM}}{6r^{5/2}}    \left[\frac{3}{r_0^3}-\frac{3}{r^3}-2\sqrt{\frac{r_c^3}{r_0^9}}+2\sqrt{\frac{r_c^3}{r_{\null}^9}}\right]~~ .
\eea
The first term on the right hand side of the equation is the usual luminosity that results from the redistribution by viscous torques of the released gravitational potential energy (Shakura \& Sunyaev 1973).  The second term represents the viscous dissipation of mechanical energy that is input to the disk via the magnetic torques.
By utilizing the definition {of eq.~(\ref{rm})}, $r_m^{7/2} = \mu^2/(\dot M \sqrt{GM})$, we can cast eq.~(\ref{lvis}) into a somewhat more aesthetically pleasing form:
{
\bea
\label{elvis}
\frac{dL_{\rm vis}}{dr} = \frac{3GM\dot M}{2r^2} \left\{ \left(1 -\sqrt{\frac{r_0}{r}} \right) + \frac{r_m^{7/2}}{9r^{1/2}r_0^3}    \left[3-3\frac{r_0^3}{r^3}-2\sqrt{\frac{r_c^3}{r_0^3}}+2\sqrt{\frac{r_c^3r_0^6}{r_{\null}^9}}\right] \right\}~~ ,
\eea
or
\bea
\label{elvisless}
\frac{dL_{\rm vis}}{dr} = \frac{3GM\dot M}{2r^2} \left\{1 +\sqrt{\frac{r_0}{r}}\left[-1 + \frac{1}{18(1-\omega)}    \left(3\omega-3\omega\frac{r_0^3}{r^3}-2 + 2\sqrt{\frac{r_0^9}{r_{\null}^9}} \right) \right]\right\}~~ ,
\eea
}

Finally, eq.~(\ref{elvis}) can be analytically integrated to yield the total power released by viscous processes in the disk:
\bea
\label{elvint}
L_{\rm vis} = \frac{GM\dot M}{2r_0} \left[1 + \frac{1}{9} \left(\frac{r_m}{r_0}\right)^{7/2} \!\!\left(4-3\sqrt{\frac{r_c^3}{r_0^3}}\right)\right] .
\eea
The first term represents the gravitational power that would be released in the disk in the absence of magnetic torques. Clearly, in this model the disk may be powered in part by the central star---this occurs  whenever $\Omega(r_0)<(4/3)\,\omega_s$, or, in terms of the fastness parameter, when 
${\omega} > 3/4$.
{In general, after the constraint of eq.~(\ref{bcfast}) is used,
eq.~(\ref{elvint}) may be rewritten as
\bea
\label{elvintless}
L_{\rm vis} = \frac{GM\dot M}{2r_0} \frac{(15-14\omega)}{18(1-\omega)}
=\frac{GM\dot M}{2r_c} \frac{(15-14\omega)}{18(1-\omega)}\omega^{-2/3}~~.
\eea
This quantity is positive for all $\omega<1$.
}

\begin{figure}[h]
\begin{minipage}[c]{0.47\textwidth}
\begin{center}
\includegraphics[height=3.0in]{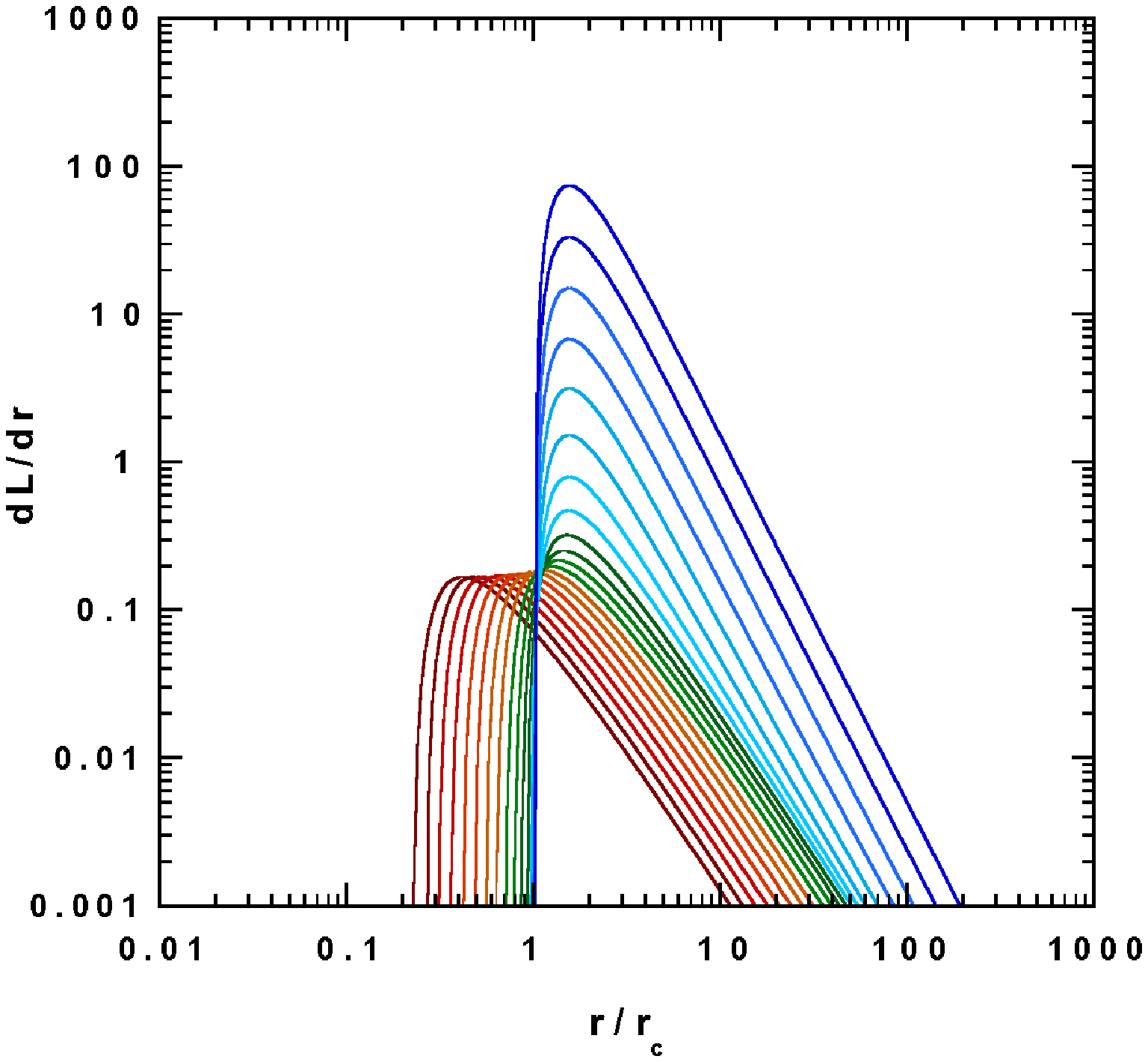} 
\caption{$dL_{\rm vis}/dr$ vs. the radial distance (in units of $r_c$) for 20 values of the parameter $\xi$ in equal logarithmic steps over the range $0.1 < \xi < 10$ (see eq.~[\ref{elvis}]); with $\xi$ increasing from left to right. 
In terms of the fastness parameter, this range of $\xi$ corresponds to $0.113 < \omega <0.99984$.
In these plots $dL_{\rm vis}/dr$ has been normalized in terms of $GM\dot M/(2r_0^2)$. 
\label{fig:dLdr1}}
\end{center}
\end{minipage}
\hfill
\begin{minipage}[c]{0.47\textwidth}
\begin{center}
\null{}\vglue0.5cm
\includegraphics[height=3.0in]{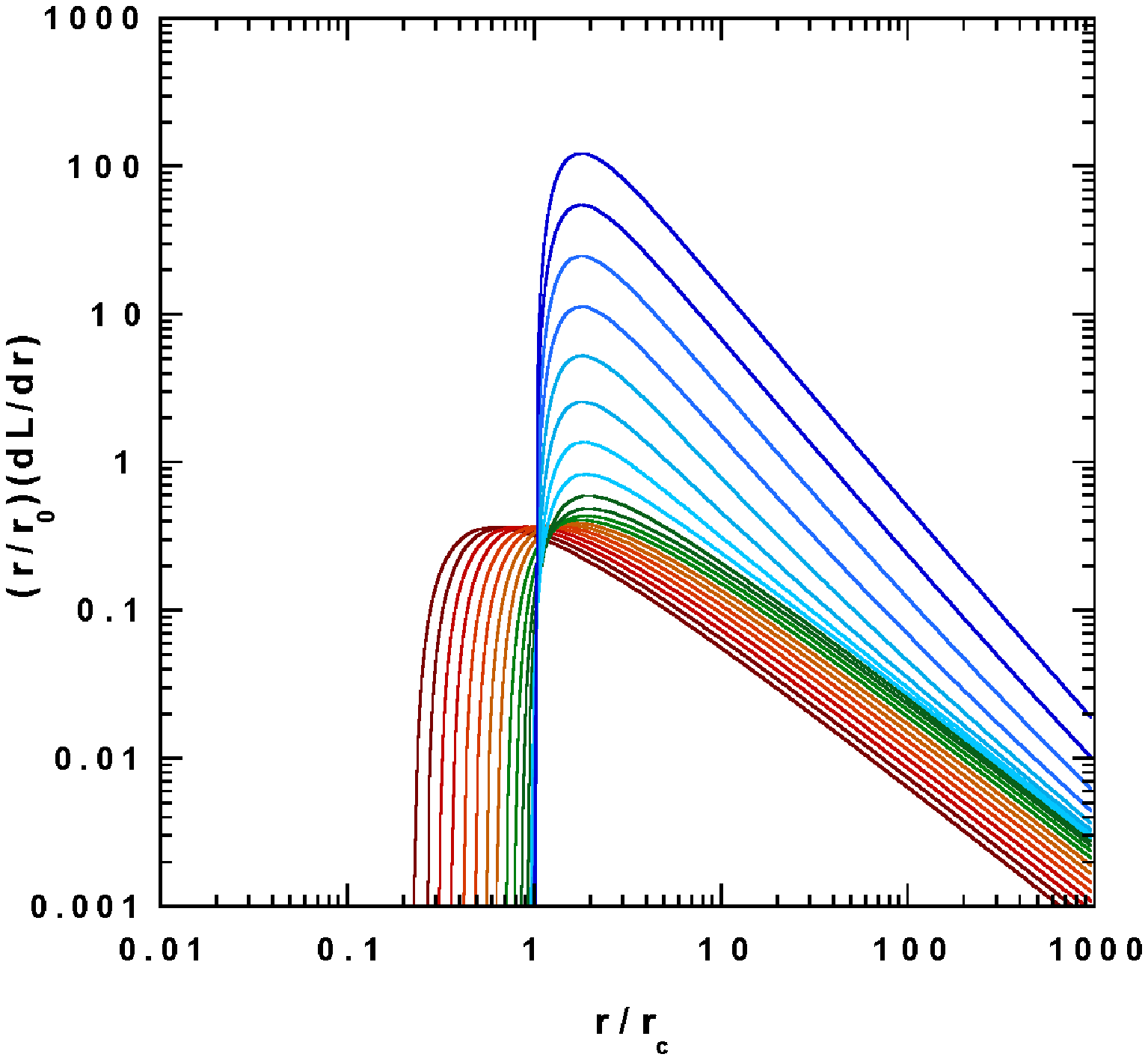}  
\caption{The quantity $(dL_{\rm vis}/dr)(r/r_0)$ vs. radial distance (in units of $r_c$) for 20 values of the parameter $\xi$ in equal logarithmic steps over the range $0.1 < \xi < 10$ (see eq.~[\ref{elvis}]); with $\xi$ increasing from left to right. In terms of the fastness parameter, this range of $\xi$ corresponds to $0.113 < \omega <0.99984$. The ordinate in these plots has been normalized in terms of $GM\dot M/(2r_0)$; the factor $r/r_0$ renders the plotted functions roughly equivalent to the luminosity per natural logarithmic radial interval. \label{fig:dPdrLn}}
\end{center}
\end{minipage}
\null{}\vglue0.25cm
\end{figure}  

The total mechanical power injected into the viscous disk and the part of the magnetosphere entraining it at $r\ge r_0$, is given by
\bea
\label{Ltota}
L_{\rm tot} = \frac{GM\dot M}{2r_0}  + \frac{\mu^2}{9r_0^3} \left(3-2\sqrt{\frac{r_c^3}{r_0^3}}\right)\omega_s~~ ,
\eea
%
where the second term on the right hand side is the rotational frequency of the central star times the total magnetic torque on the disk (see eq.~[\ref{taub}] below).
{For $\dot M\ne 0$,} this expression reduces to:
\bea
\label{Ltot}
L_{\rm tot} = \frac{GM\dot M}{2r_0} \left[1 + \frac{2}{9} \left(\frac{r_m}{r_0}\right)^{7/2} \!\!\left(3\sqrt{\frac{r_0^3}{r_c^3}}-2\right)\right].
\eea
{Eq.~(\ref{Ltot}) can be rewritten as
\bea
\label{Ltotless}
L_{\rm tot}=\frac{GM\dot M}{2r_0}\left[1 + \frac{1}{9}\frac{\omega (3\omega-2)}{(1-\omega)}\right]
=\frac{GM\dot M}{2r_c}\left[1 + \frac{1}{9}\frac{\omega (3\omega-2)}{(1-\omega)}\right]\omega^{-2/3}~.
\eea
The difference between the expressions in eq.~(\ref{Ltotless}) and eq.~(\ref{elvintless}) is
\bea
\label{Lmismatch}
L_{\rm tot} - L_{\rm vis}=\left(\frac{GM\dot M}{2r_c}\right) \frac{\omega^{-2/3}\left(3-8\omega+6\omega^2\!\right)}{18(1-\omega)}~~,
\eea
a positive quantity for all $\omega<1$.
}

It is not clear what fraction of the difference $L_{\rm tot} - L_{\rm vis}$
is released in the disk, and what fraction is released in the magnetosphere
(perhaps as non-thermal radiation).
  In the limit of $\dot M \rightarrow 0$, $L_{\rm tot} = 2L_{\rm vis}$, and $L_{\rm vis}=\omega_s \tau_c/2$, where $\tau_c= \mu^2/(9r_c^3)$.


\subsection{Magnetic torques on the disk and the neutron star}
\label{sec:torques}

The total magnetic torque on the accretion disk, $\tau_{B,{\rm disk}}$, is given by the magnetic contribution to $- 2\pi r^2 T_{r,\phi}$ evaluated as $r \rightarrow \infty$ (this is equivalent to the integral on the right hand side of eq.~[\ref{govern}] integrated to $\infty$).  This quantity can be found from the second term on the right side of eq.~(\ref{visc2}), multiplied by $2\pi r^2$, and in the limit  $r \rightarrow \infty$:
\bea
\label{taub}
 \tau_{B,{\rm disk}}= \frac{\mu^2}{9r_0^3}  \left[3-2\sqrt{\frac{r_c^3}{r_0^3}}\right] ~.
\eea
{For $\dot M\ne 0$ eq.~(\ref{taub}) may be rewritten in terms of the fastness (eq.~[\ref{fastness}]):
\bea
\label{taubfast}
\tau_{B,{\rm disk}}= \left(\dot M\sqrt{GMr_0}\right)\,\frac{(3\omega-2)}{18(1-\omega)} ~~.
\eea
}
It may be interesting to note that the net torque on the disk vanishes, i.e., $\tau_{B,{\rm disk}}=0$, when $\Omega(r_0)=(3/2)\omega_s$.
In the limit of $\dot M \rightarrow 0$, $ \tau_{B,{\rm disk}}\rightarrow \tau_c= \mu^2/(9r_c^3) = \dot M_0\sqrt{GMr_c}/9$.

{ Here, we can compare our results with the numerical work of Agapitou \& Papaloizou (2000). If the disk extended only between $r_0=r_c$ and $r=r_{\rm ext}$, as Agapitou \& Papaloizou (2000) assume in their calculations, we would have obtained
$$\tau_1=(\mu^2/3r_c^3)\left[1/3 -r_c^3/r_{\rm ext}^3 +(2/3)(r_c/r_{\rm ext})^{9/2}\right]$$
instead of eq.~(\ref{taub}). Agapitou \& Papaloizou take $B_\phi = C B_\phi^I$ (see our eq.~[\ref{bphi}]) with $B_z$ found numerically to be reduced relative to the dipole value that we assume. The torque $N(C)$ they find in their numerical solution is a few times larger than $\tau_1$, with $N(C)/\tau_1\sim 2$ for $C\sim100$---see eq.~[26] and Fig.~14 of Agapitou \& Papaloizou (2000). Thus, our simple model yields results for the torque which are not too different from those found for more realistic magnetic field configurations. 
Since in reality $r_0\ne r_c$, we note that rescaling of the product of the toroidal and poloidal magnetic fields, $B_\phi B_z$, by a constant factor, $C_1$, affects our results only to the extent that $r_0$ varies, so that the values of $\omega$ are rescaled (eqs.~[\ref{bc}], [\ref{bcfast}]). At the same time the constraint of eq.~(\ref{bc}) guarantees that there is no explicit dependence on $C_1$ in eq.~(\ref{taubfast}), and in many of our other equations. Formally, $r_0$ and $\tau_{B,{\rm disk}}$ are invariant under the rescaling $B_\phi B_z\rightarrow C_1B_\phi B_z$ and $\dot M\rightarrow C_1^2\dot M$, see eq.~(\ref{rm}).
}

The magnetic torque on the central star is just the negative of eq.~(\ref{taub}).  To find the total torque on the central star we add the material torque:
\bea
\label{taus}
 \tau_{\rm CS}= \dot M \left(\sqrt{GMr_0} - \omega_s R_s^2 \right) -\frac{\mu^2}{9r_0^3}  \left[3-2\sqrt{\frac{r_c^3}{r_0^3}}\right],~~\nonumber\\
 \eea
where $R_s$ is the radius of the central star. (Note that $\omega_s R_s^2$ is taken to be negligible in the production of our graphical results.)
{For high accretion rates, the material torque is the leading term in eq.~(\ref{taus}),  $\tau_{\rm CS}\sim \dot M \sqrt{GMr_0}\propto \dot M^{9/10}$.
}

If we combine eqs.~(\ref{taus}),
{
 (\ref{rm}), and (\ref{bcfast}),
}
we can find an analytic expression for the condition that there is zero torque on the central star (under the approximation that we neglect the $\omega_sR_s^2$ term in eq.~[\ref{taus}], 
{
cf., eq.~[\ref{dimlesstorque}]).
}
We find 
{$r_0/r_c=(20/21)^{2/3}$, and $r_m/r_0=10^{2/7}$, yielding}
$\xi_{\tau = 0} = 1.869$ and $r_0^{\tau=0} =0.9680 r_c = 0.5179 r_m^{\tau=0}$.  The corresponding values for 
{our alternate model (Appendix~\ref{AppB}) are very similar:}
$\xi_{\tau = 0} = 1.869$ and $r_0^{\tau=0} =0.9666 r_c = 0.5172 r_m^{\tau=0}$.

\begin{figure}[t]
\begin{minipage}[c]{0.47\textwidth}
\begin{center}
\includegraphics[height=3.0in]{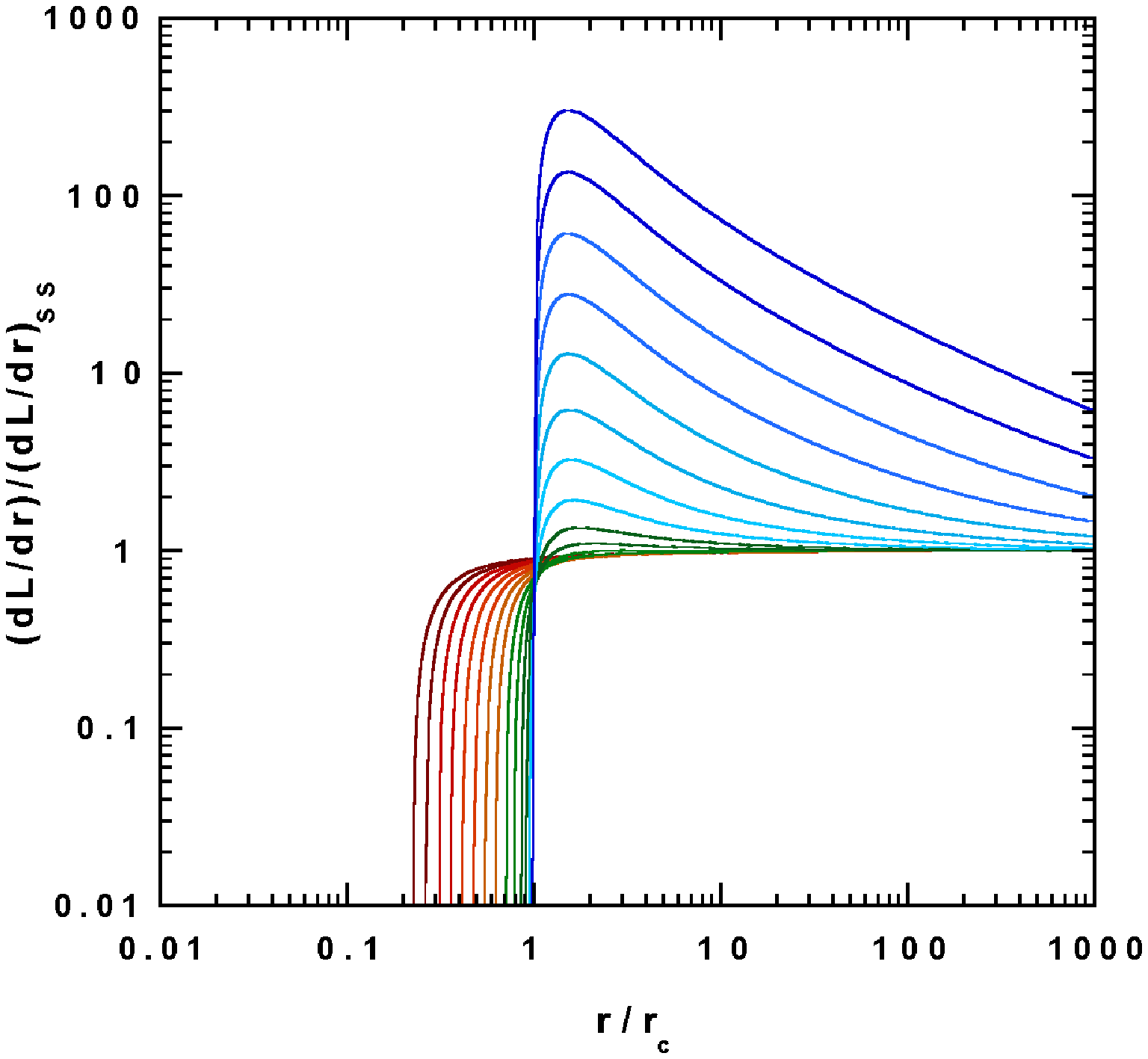} 
\caption{The quantity $dL_{\rm vis}/dr$ vs. the radial distance (in units of $r_c$) for 20 values of the parameter $\xi$ in equal logarithmic steps over the range $0.1 < \xi < 10$ (see eq. [\ref{elvis}]); with $\xi$ increasing from left to right. In terms of the fastness parameter, this range of $\xi$ corresponds to $0.113 < \omega <0.99984$. These plots have been normalized to $dL_{\rm vis}/dr$ for a Shakura-Sunyaev disk with the same values of $\dot M$ and inner radius, $r_0$. 
\label{fig:dPdrSS}}
\end{center}
\end{minipage}
\hfill
\begin{minipage}[c]{0.47\textwidth}
\begin{center}
\null{}\vglue-1.6cm
\includegraphics[height=3.0in]{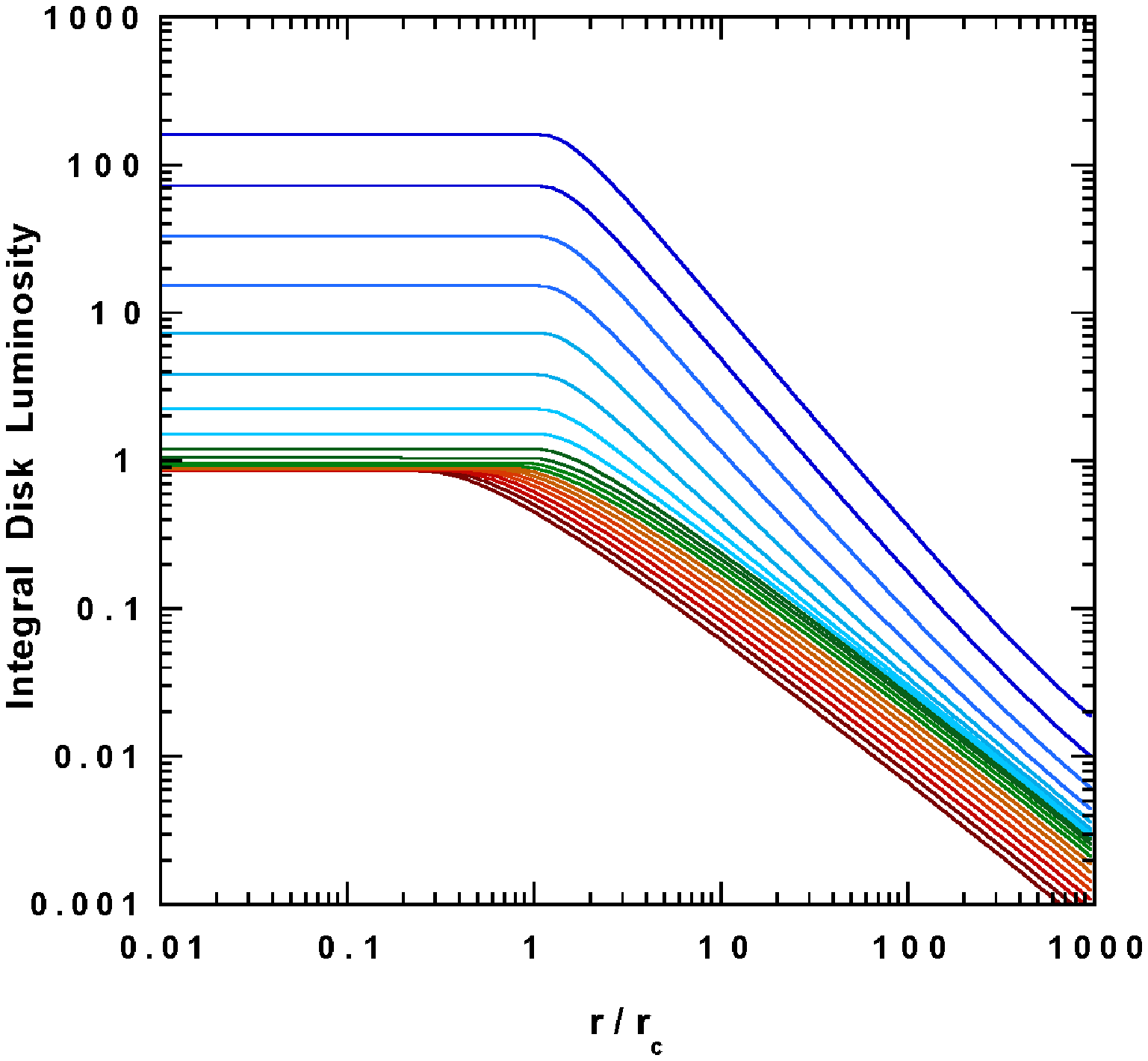} 
\caption{The integral curves for $dL_{\rm vis}/dr$ as given by Fig. 3. \label{fig:dPdrInt}}
\end{center}
\end{minipage}
\end{figure}  
\clearpage

\section{Results in Graphical Form}
\label{sec:results}

The inner disk transition radii, $r_0$, as given by eqs.~(\ref{bc}), (\ref{bcbis}), are shown in Fig.~\ref{fig:rad}.  The four plotted curves are $r_0$ in units of $r_m$, and $r_0$ in units of $r_c$, for both of our prescriptions for $B_\phi$.  These radii are plotted as a function of the parameter $\xi^{-1} = r_c/r_m$. Some fiducial dimensionless values of $r_0$ and $r_m$ can be found in Table~\ref{table1}.

\begin{table*}
\centering
\caption{{Some fiducial radii}}
\centering
\label{table1}
\begin{tabular}{cccccc}

\hline
\hline
~~~~~eq.~(\ref{fastness})~~~~ & & ~~~~~eq.~(\ref{bc})~~~~ & & ~~~~~eq.~(\ref{dotM0})~~~~ & \\ 
\hline
\hline
$\omega$ & $r_0/r_c$ & $\xi$ & $\xi^{-1}$ & $\dot M/\dot M_0$ & comment\\
\hline
\hline
0.9524& 0.9680 & 1.869~  & 0.5351 & 0.1120 & $\tau_{CS}=0$\\
0.7808& 0.8478 & 1      & 1      & 1      & definition of $\dot M_0$ \\
3/4   & 0.8255 & 0.9269 & 1.079  & 1.304  & $L_{\rm vis}=GM\dot M/(2r_0)$ \\  
2/3   & 0.7631 & 0.7634 & 1.310  & 2.576  & $\tau_B=0$, $L_{\rm tot}=GM\dot M/(2r_0)$\\

\hline
\hline
\end{tabular}
\vskip0.5truecm
\end{table*}%

 In the slow rotator limit (i.e., weak $B$ or high $\dot M$; large $\xi^{-1}$), $r_0/r_c \rightarrow 0$, as expected, while $r_0/r_m$ asymptotically approaches $2^{2/7}=1.22$ for the $B^{II}_\phi$ model, and $r_0/r_m$ is larger for the $B^{I}_\phi$ model, slowly increasing with increasing $\xi^{-1}$.
In the opposite limit of a rapid rotator (i.e., strong B or low $\dot M$; small $\xi^{-1}$), we see that $r_0 \rightarrow r_c$  as $\xi^{-1} \rightarrow 0$, as discussed extensively by Rappaport et al. (2004).  In that paper, the authors suggest that for reasonably fast central rotators, the matter is not ejected via a propeller mechanism, but rather the accretion disk adjusts its structure to allow matter to reach $r_c$ (see also \S4.2 in this article). Note that for this and other reasons, $r_0/r_m$ varies quite strongly with $r_c/r_m$, in contrast with the Ghosh \& Lamb (1979) model, where the same ratio, $r_0/r_m$, is a constant equal to about 0.4. In particular, we find that $r_0/r_m \rightarrow 0$ as $\xi^{-1} \rightarrow 0$.

In Figures 3 through 5 we plot the viscous luminosity in the disk per radial interval, $dL_{\rm vis}/dr$, eq.~(\ref{elvis}), with three different normalizations.  In the first of these (Fig. 3), $dL_{\rm vis}/dr$ is normalized to the differential gravitational luminosity at $r_0$, viz, $GM\dot M/(2r_0^2)$.  Fig. 4 is $dL_{\rm vis}/dr$ multiplied by a factor of $r/r_0$ to yield the viscous luminosity per natural logarithmic radial interval.  In Fig. 5, $dL_{\rm vis}/dr$ is normalized to the same quantity for a Shakura \& Sunyaev disk.  In all cases, the pattern of the curves indicates how the inner radius of the disk moves outward as the parameter $\xi$ increases (i.e., as the central object becomes a faster rotator), and ultimately asymptotically approaches $r_c$.  For very rapid rotators, the plots clearly indicate that the disk energetics are dominated by the mechanical energy being pumped in by the magnetic field, and dissipated (non-locally) by viscous stresses.  For the rapid rotators, the extra energy being pumped into the disk is powered by a spin-down of the central star.  Finally, in Fig. 6 we show the integral viscous disk luminosity profiles, i.e., $L_{\rm vis}$ released for radii $>r$.

We note that in Figs. 3--6, due to the way the curves have been normalized, the large disk luminosities for rapid rotators are {\em only with respect to $L_{\rm vis}$ in the absence of magnetic torques}.  The absolute values would steadily {\em decrease} as $\xi$ grows, since $\dot M$ is decreasing in the normalization factor $GM\dot M/(2r_0)$.  
The absolute value of the disk luminosity (i.e., not normalized to $1/r_0$) is shown in Fig. 7 as a function of $\dot M$.  In order to make these plots as generic as possible we have normalized the luminosities to $GM\dot M_0/(2r_c)$, where $r_c$ is fixed for a given system, and $\dot M$ is normalized to $\dot M_0 \equiv \mu^2/(r_c^{7/2}\sqrt{GM})$, i.e., a value that is also fixed for a given system.  The three luminosities shown in the figure are (i) $L_{\rm vis}$, as given by eq.~(\ref{elvint}), and which includes the mechanical energies put into the disk via the gravitational field and the magnetic torques; (ii) $L_{\rm \dot M} \equiv GM\dot M/(2r_0)$ which results from the deposited gravitational energy alone; and (iii) the total energy released in the system, including that due to an equivalent ``frictional" dissipation via the magnetic field (see eq.~[\ref{Ltot}]). The effects of magnetic heating of the disk are very apparent for low values of $\dot M$, i.e., the fast rotator case.  The effects of magnetic ``cooling'', for slow rotators, are much less apparent.

The torques on the central star are shown in Fig. 8.  The four curves are combinations of $|\tau_{\rm B}|$ and $|\tau_{\rm tot}|$ vs. $L_{\rm vis}$ or $L_{\rm tot}$.  Here $L_{\rm vis}$ and $L_{\rm tot}$ are defined by eqs.~(\ref{elvint}) and (\ref{Ltot}), respectively.  The torques, $\tau_{\rm B}$ and $\tau_{\rm tot}$, are defined in eqs.~(\ref{taub}) and (\ref{taus}), respectively, where $\tau_{\rm tot} \equiv \tau_{CS}+\dot M\omega_s R_s^2$.  For high values of the disk luminosity we see that the magnetic torques are typically an order of magnitude smaller than the matter accretion torques.  At lower luminosities, the magnetic torque switches sign at a luminosity of about (1.2 - 1.3)\,$GM\dot M/(r_c)$, while the total torque (including that due to accretion) reverses sign only at luminosities about an order of magnitude lower (cf. Table~\ref{table1}).  Note that the negative torques tend to a constant value, as the disk luminosity decreases (i.e., as the mass accretion rate drops).

In terms of the spin-up/spin-down dependence on the mass accretion rate, our results are qualitatively similar to those of Wang 1987, 1996, and quite different from those of Ghosh \& Lamb (1979). With the help of eq.~(\ref{bc}), and neglecting the term $\omega_s R_s^2$, the torque of eq.~(\ref{taus}) can be expressed as a dimensionless function $g_I$ of the fastness parameter:
{
\bea  
 \label{dimlesstorqueIalt}
\tau_{\rm CS}=\left(\dot M \sqrt{GMr_0}\right)\,g_I(\omega),
\eea
or, with a different normalization,
\bea  
 \label{dimlesstorqueI}
\tau_{\rm CS}=\left( \dot M \sqrt{GMr_c}\right) \omega^{1/3}\,g_I(\omega),
\eea
with
\bea
\label{dimlesstorque}
g_I(\omega)={10\over 9}\left[1 -{1\over 20} {\omega\over 1- \omega}\right].
\eea
}
As in Wang's (1996) dimensionless torque, $g_I$ is a fairly flat function up to the zero at $\omega=20/21=0.9524\approx (0.9680)^{3/2}$, in general agreement with the observations of Finger et al. (1996) and in contrast to Ghosh \& Lamb's (1979) steeply decreasing torque which vanishes at $\omega\approx0.35$. 
{Remarkably, the corresponding function $g_{II}$ in our alternate model, although also fairly flat with a value higher by a factor $\simeq 63/60$, has a first zero at a very similar value of $\omega =  0.9502$ (eq.~[\ref{dimlesstorqueIIb}]).
}

 As discussed
{in \S~\ref{sec:torques}}
above, the condition for zero torque on the central star is $\xi_{\tau = 0} = 1.869$ ($\xi_{\tau = 0}^{-1} = 0.535$) for {both of our models. This corresponds} to $r_0/r_m \simeq 0.52$, {a value which coincidentally happens to be} in approximate agreement with the Ghosh \& Lamb (1979) model
{prediction of $r_0/r_m \simeq 0.42$ for any torque (Fig.~\ref{fig:rad}). However, the value of fastness at this point, $\omega_{\tau = 0}=0.95$ is quite different from the Ghosh \& Lamb (1979) value of $\omega\approx0.35$.
}

The zero torque constraint on $\xi$ leads to a relation between the equilibrium spin of a neutron star at fixed $\dot M$ and the surface $B$ field:
$P_s \simeq 2.9 \xi_{\tau = 0}^{-3/2} \mu_{27}^{6/7} \dot M_{18}^{-3/7}~~{\rm ms}$, where $\mu_{27}$ is the neutron star magnetic moment in units of $10^{27}$ G cm$^{-3}$, and $\dot M_{18}$ is the mass accretion rate in units of $10^{18}$ g s$^{-1}$, or approximately the Eddington limit for a neutron star.  The equilibrium spin period turns out to be 
{
\bea
\label{spinup}
P_s = 1.14 ~{\rm ms} ~(\dot M_{18})^{-3/7} B_9^{6/7}~~,
\eea
where $B_9$ is the surface value of the magnetic field in units of $10^{9}$ G.}
Unfortunately, at present it is not clear how close to equilibrium the various X-ray pulsars are.

Observations of X-ray pulsars indicate a bimodal behavior of spin-up and spin-down, with a rapid change between two states of similar $|\dot P|$ magnitude, but of opposite sign of the rate of change of the pulsar spin, $\dot P$ (Bildsten et al. 1997). The change between the two states can be quite abrupt, and occurs without a large change in (pulsed) luminosity. In our model,  a change in luminosity over a fairly narrow range of luminosity could, indeed, lead to a change from spin-up torque to a spin-down torque of  comparable magnitude (or vice versa), see Fig.~8. Further, the spin-down torque quickly tends to a constant value, as the luminosity drops. This is in rough qualitative agreement with the observations, however the model seems to admit a wider range of torques, particularly of spin-up torques, than is oberved. In particular, recent observations suggest very little change, if any, of the magnitude of the torque during the observed transition from spin-up to spin-down in 4U 1626-67 (Krauss et al. 2007, see their Fig.~5).

\begin{figure}[t]
\begin{minipage}[c]{0.47\textwidth}
\begin{center}
\includegraphics[height=3.1in]{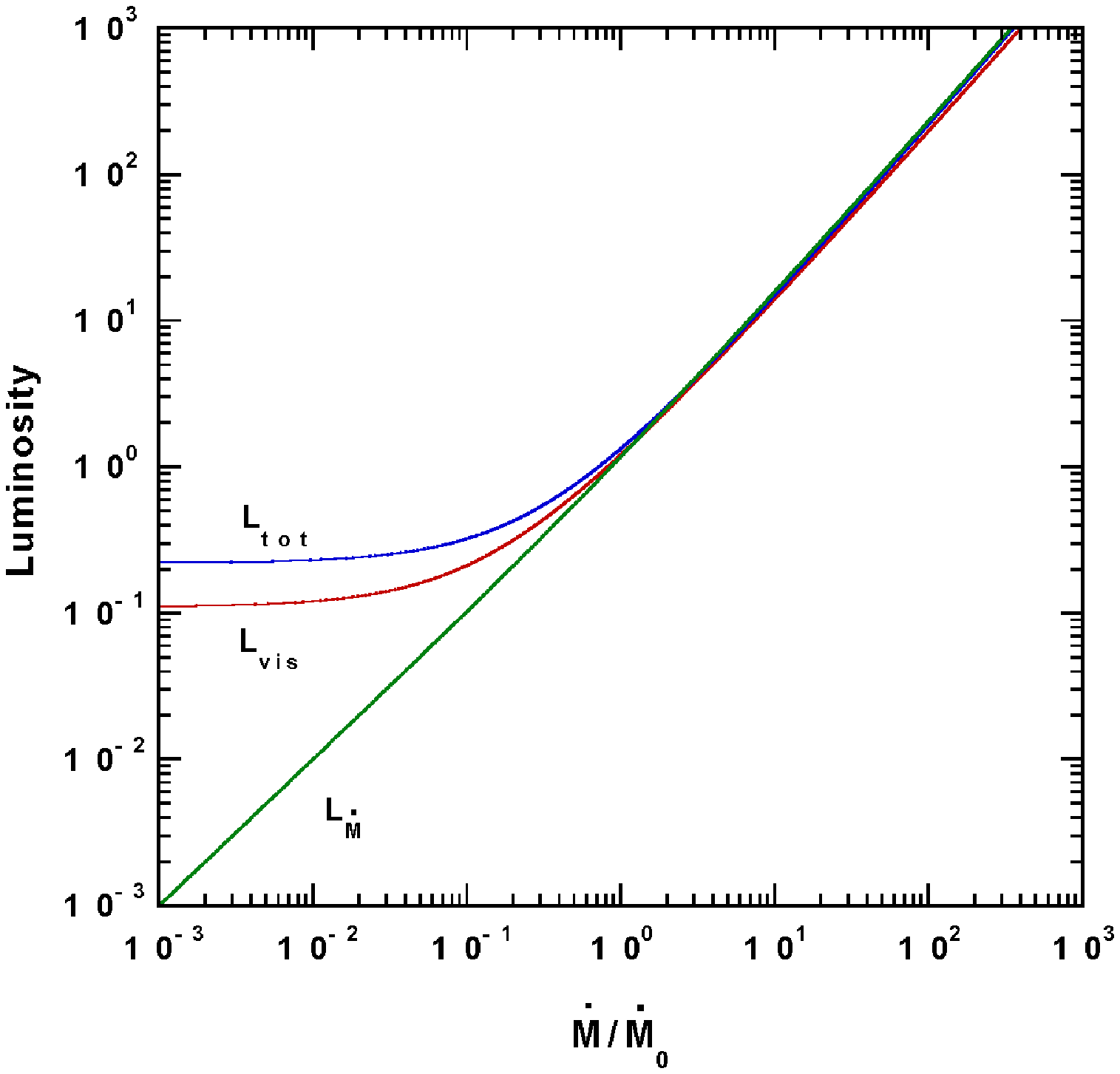}  
\caption{Luminosity as a function of $\dot M$. The reference mass
transfer rate, $\dot M_0$, is the value of $\dot M$ which, for given 
values of $\mu$, $M$, and $r_c$, yields a $\xi$-parameter value of unity.  
The curve labeled $L_{\rm vis}$ describes the total viscous luminosity of the disk, 
as given by eq.~(\ref{elvint}), while the curve labeled $L_{\rm tot}$ is the total 
mechanical energy put into the system via the magnetic field and gravity -- 
including the magnetic ``frictional'' dissipation term (eq.~[\ref{Ltot}]).  The curve labeled $L_{\dot M}$ is the gravitational energy release, $GM\dot M/(2r_0)$.  All luminosities are expressed in units of $GM\dot M_0/(2r_c)$.
{In this paper, we do not discuss the contribution to luminosity of the power released by matter in transit from the viscous disk to the stellar surface.
}
\label{fig:lum}}
\end{center}
\end{minipage}
\hfill
\begin{minipage}[c]{0.47\textwidth}
\begin{center}
\null{}\vglue-1.2cm
\includegraphics[height=3.1in]{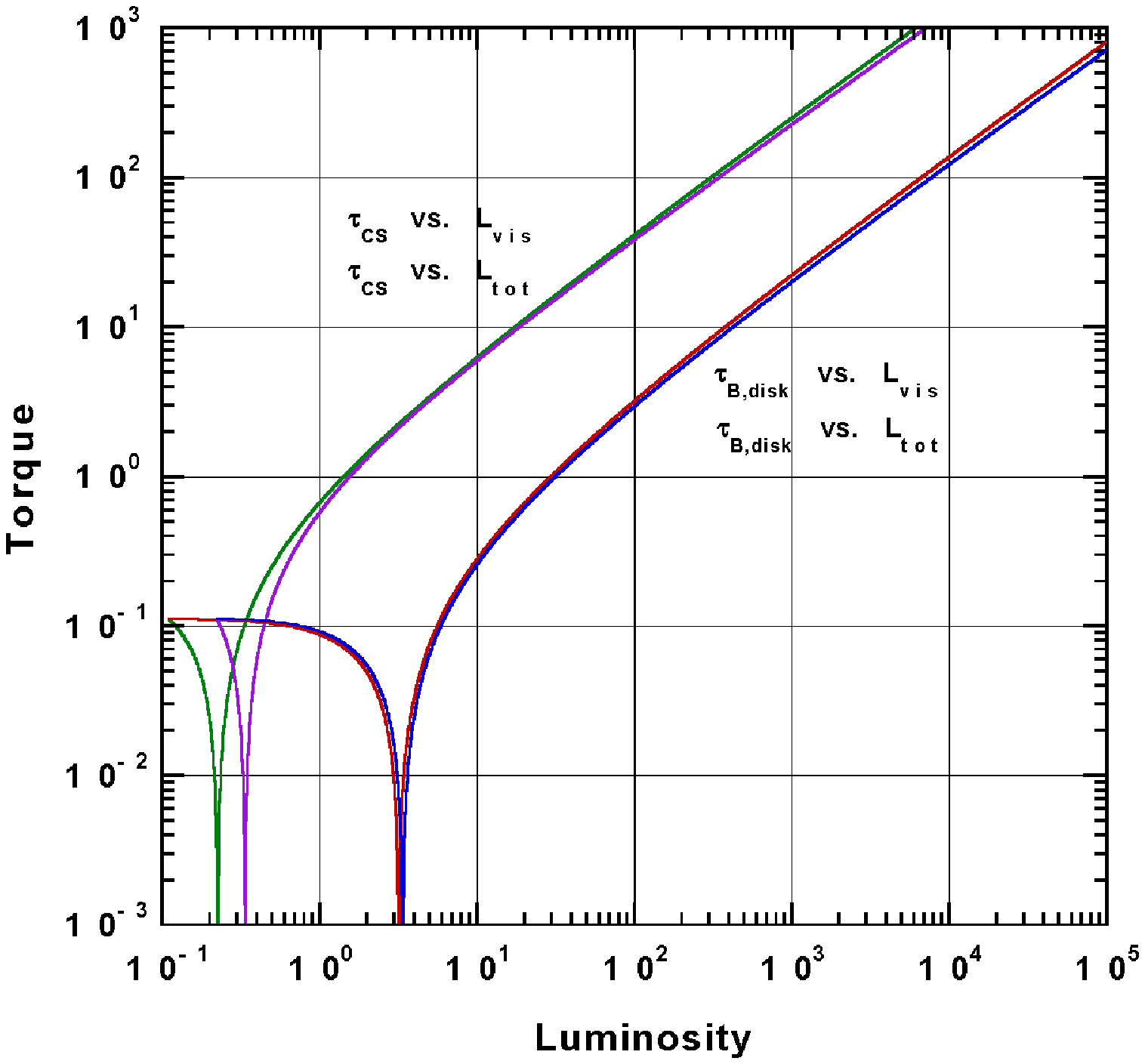}
\caption{The torque on the central star due to magnetic torques (right set of curves, negative of eq.~[\ref{taub}]) and magnetic plus matter torques (left set, eq.~[\ref{taus}]), vs. $L_{\rm vis}$ and $L_{\rm tot}$ (eqs.~[\ref{elvint}], [\ref{Ltot}]). 
The $\dot M$ dependence of these luminosities is shown in Fig. 7.  The torques and luminosities
are normalized to $\dot M_0 \sqrt{GMr_c}$ and $GM\dot M_0/(2r_c)$, respectively,
where $\dot M_0$ is defined in the caption to Fig. 7.  Only the absolute magnitudes of the torques are plotted to allow for the use of a log scale.
\label{fig:torques}}
\null{}\vglue0.6cm
\end{center}
\end{minipage}
\end{figure}  

\section{Complete Thin Alpha-Disk Solutions}
\label{sec:disksoln}

Our expression for the vertically integrated viscous stress (eq.~[\ref{visc2}]) can be used to derive the full set of thin accretion disk variables if we specify a prescription for the stress.  The most often utilized of these is the ``alpha-prescription'' of the Shakura-Sunyaev model (1973) in which $|T_{r,\phi}|$ is taken to be $\alpha P \cdot 2H$, where $P$ is the disk pressure in the midplane, and $\alpha$ is a dimensionless parameter which specifies the strength of the viscous forces.  Equation~(\ref{visc2}) then directly yields:
\bea
\label{peh}
2PH =  \frac{\dot M \Omega_K}{2 \pi \alpha} \times 
{
F(\sqrt{r_0/r},\omega)
} ~~,
\eea
{with}
\bea
\label{F}
F =  \left(1 -\sqrt{\frac{r_0}{r}} \right) + \frac{1}{9} \left(\frac{r_m}{r_0}\right)^{7/2} \!\!\!\! \sqrt{\frac{r_0}{r}}   \left[3-\frac{3r_0^3}{r^3}-2\sqrt{\frac{r_c^3}{r_0^3}}+2\sqrt{\frac{r_c^3r_0^6}{r_{\null}^9}}\right]~~.
\eea
Here, the first term in parentheses in the expression for $F$ is the usual factor $f(r)^4$ in the Shakura-Sunyaev solution.
{If it were not for the constraint of eq.~(\ref{bc}),  putting $r_m=0$ one would recover the Shakura-Sunyaev solution, but this would be inconsistent as $r_0$ is no longer an arbitrary constant, and instead must tend to zero with $r_m$.
After eliminating $r_m$ from eq.~(\ref{F}) with the help of eq.~(\ref{bcfast}), we obtain
\bea
\label{Ffast}
F(y,\omega) = [18(1-\omega)]^{-1}\left(18 - 20y -18\omega + 21\omega y - 3\omega y^7 + 2 y^{10}\right)~,
\eea
in the domain $0<y\le 1$, $0 \le \omega <1$. Here, $y\equiv\sqrt{r_0/r}$, and $\omega$ is defined in eq.~(\ref{fastness}). The limit of vanishing field, $\mu\rightarrow 0$ is given by
\bea
\label{Flimit}
F(y,0)=(1-y) - \left(y-y^{10}\,\right)/9,
\eea
because $\omega=0$ in this limit (eq.~[\ref{bc}]). For a fixed value of $r$ (or one bounded from below), $y\rightarrow0$, $F\rightarrow1$, in the same limit.
}

Three of the four remaining SS-disk equations remain unchanged.  For the equation of state we take: $P=\rho k T/m$, where $m$ is the mean molecular weight per particle.  The vertical force balance equation is given by: $P \simeq GM\rho H^2 r^{-3}$.  Radiative transport in the vertical direction is represented by: $T^4 \simeq \kappa \rho H T_e^4$, where $T$ and $T_e$ are the midplane and effective temperatures of the disk, respectively, and $\kappa$ is the radiative opacity evaluated at the disk midplane.  Finally, we retain the fifth SS-disk equation for the heat dissipation per unit surface area of the disk: $\mathcal{F} =\dot QH \simeq \alpha H P \Omega = \sigma T_e^4$.  In doing so, we have explicitly neglected the magnetic ``frictional'' heating; this latter heat input is never dominant over the mechanical energy deposited by the magnetic field.  For this reason, and in order to keep the disk equations in strictly algebraically solvable form, we neglect the magnetic ``frictional'' heating.  In all of these equations (except for the equation of state) we have neglected dimensionless coefficients of order unity.  These factors have no impact on the form of the solutions for $P$, $\rho$, $H$, and $T$, and only a minor effect on the leading coefficients to the solutions.

The four equations listed above, plus the equation for $PH$ given in equation (\ref{peh}) can be solved algebraically in a manner analogous to that done for the original SS-disk equations to yield $P$, $\rho$, $H$, and $T$ as functions of $r$, $\dot M$, $\alpha$, and $\mu$.  In solving these equations we have taken $\kappa$ to be given by Kramers opacity ($\kappa \simeq 6 \times 10^{22} \rho T^{-3.5}$ cm$^2$ gm$^{-1}$) which is appropriate for most of the physical conditions found in our disk models, and the mass of the central star is fixed at $1.5~M_\odot$.  The solutions are:
\bea
\label{P}
P \simeq 2 \times 10^5 \alpha^{-9/10} \dot M_{\rm 16}^{17/20} r_{\rm 10}^{-21/8} F^{17/20}
~{\rm dynes\,cm^{-2}}
\eea
\bea
\label{twoh}
H \simeq 0.5 \times 10^8 \alpha^{-1/10} \dot M_{\rm 16}^{3/20} r_{\rm 10}^{9/8} F^{3/20}
~~~~~{\rm cm}
\eea
\bea
\label{Temp}
T \simeq 2 \times 10^4 \alpha^{-1/5} \dot M_{\rm 16}^{3/10} r_{\rm 10}^{-3/4} F^{3/10}
~~~~~{\rm K}
\eea
\bea
\label{rho}
\rho \simeq 7 \times 10^{-8} \alpha^{-7/10} \dot M_{\rm 16}^{11/20} r_{\rm 10}^{-15/8} F^{11/20}
~~~~~{\rm g~cm}^{-3}
\eea
where $\dot M_{\rm 16}$ is the mass accretion rate in units of $10^{16}$ gm sec$^{-1}$, and $r_{\rm 10}$ is the radial distance in units of $10^{10}$ cm. 
{ Analogous expressions have been derived by Matthews et al. (2005); see eq.~(\ref{FII}) in our Appendix B.
}
{Note that at $r=r_0$ for any magnetic field value $F(1,\omega) =  f(r_0) = 0$. Thus, our expressions for $P$, $H$, $T$, $\rho$ (eqs.~[\ref{P}, \ref{twoh}, \ref{Temp}, \ref{rho}] will vanish at $r=r_0$, just as in the Shakura-Sunyaev (1973) solution. This non-physical result is an artifact of the alpha viscosity prescription.
 Therefore, eqs.~(\ref{P}, \ref{twoh}, \ref{Temp}, \ref{rho}) are not expected to be an accurate description of the disk close to $r\approx r_0$.
}
 
Eqs.~(\ref{P})--(\ref{rho}) are derived under the assumptions that the accretion disks are optically thick, the radiative opacities are given by Kramers opacity, and gas pressure dominates over radiation pressure.  The latter two assumptions can be changed to electron scattering opacity and/or a dominant radiation pressure, while still retaining analytic solutions.  Some of these solutions, especially for values of $\dot M$ approaching Eddington, may be thermally unstable (e.g., the Lightman-Eardley instability; 1974), but this subject is beyond the scope of the present paper.

\clearpage
\begin{figure}[t]
\begin{minipage}[c]{0.94\textwidth}
\begin{center}
\includegraphics[height=3.0in]{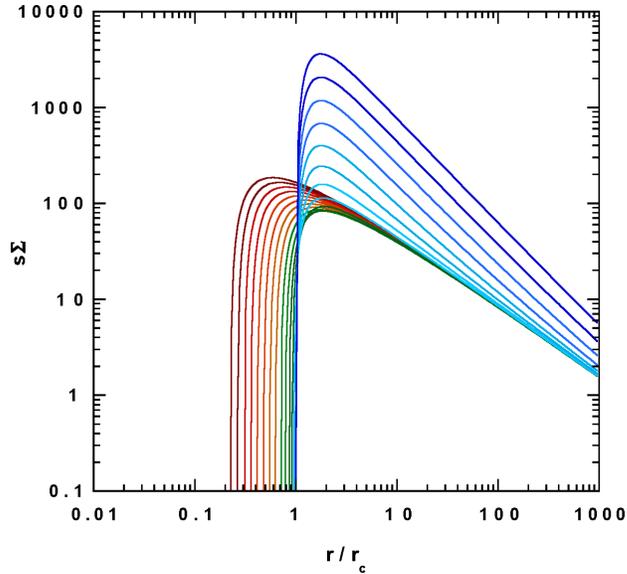}
\caption{Surface density, $\Sigma$, vs. the radial distance (in units of $r_c$) for 20 values of the parameter $\xi$ in equal logarithmic steps over the range $0.1 < \xi < 10$; with $\xi$ increasing from left to right. $\Sigma$, in units of g cm$^{-3}$, has been computed as the product of $\rho$ and $2H$, from eqs.~(\ref{rho}) and (\ref{twoh}), respectively.  The value of $\dot M$ used to evaluate $\Sigma(r)$ has been arbitrarily fixed at $10^{16}$ g s$^{-1}$, but the curves scale simply as $\dot M^{7/10}$.  The multiplicative constant ``$s$'' that multiplies $\Sigma$ is defined as $r_{c,10}^{3/4}$; this is done to allow the radial coordinate to be scaled in units of $r_c$.
\label{fig:sigma}}
\end{center}
\end{minipage}
\hglue0.05cm
\end{figure}

In practice these equations could be utilized in the following way.  For any accreting system which is thought to have a thin accretion disk, the accretion rate, and the mass, rotation rate, and magnetic moment of the central star would be combined to evaluate the parameter $\xi \equiv \left[\mu^2/(\sqrt{GM} \dot M)\right]^{2/7}\!\!r_c^{-1}$
{
(see eqs.~[\ref{rc}], [\ref{rm}], [\ref{xi}]).  This value of $\xi$ would be used in eq.~(\ref{bc}) (or eq.~[\ref{bcbis}], depending on the model for $B_\phi$), to solve for $r_0$, and for $\omega$ (eqs.~[\ref{bcfast}], [\ref{bcIIfast}]).  With $r_0$ and $\omega$ in hand, eq.~(\ref{Ffast}) (or eqs.~[\ref{FIIfast}] and [\ref{FIIbfast}]) yields the function $F(\sqrt{r_0/r},\omega)$, 
}
which is required to solve for the pressure, density, and temperature in the disk via eqs.~(\ref{P})--(\ref{rho}).

An illustrative example of disk radial profiles is shown as surface density plots in Fig.\,9.  Here we show $\Sigma(r)$ for a range of values of the parameter $\xi$.  The plots shown are for an arbitrary fixed value of $\dot M =10^{16}$ g s$^{-1}$, but the curves scale simply as $\dot M^{7/10}$.  The multiplicative constant ``$s$'' that multiplies $\Sigma$ is defined as $r_{c,10}^{3/4}$, with $r_{c,10}\equiv r_c/(10^{10}\,{\rm cm})$; this is done to allow the radial coordinate to be scaled in units of $r_c$.
{
In essence, up to a normalization factor, these are plots of $y^{3/2} F^{7/10}(y,\omega)$ versus $\omega^{2/3}y^{-2}$, with each curve representing a different fixed value of $\omega$ (or, equivalently, $\chi$).
}

\section{Summary and Conclusions}
\label{sec:sum}

We have considered an analytic model of interaction between an accretion disk and a rotating magnetic field of the central star, in which the magnetic torques are distributed over a range of radii in the inner disk. Neglecting mass loss from the disk, we have solved the disk angular momentum equation to obtain a radial angular velocity profile smoothly matching the angular velocity of the star to that of the Keplerian disk.

We have solved self-consistently for the inner  radius of the
{\em viscous}
disk, and have found that it is within the corotation radius, $r_0<r_c$ (\S4).
{The magnetically dominated disk extends further in, with the maximum value of $\Omega$ attained at $r_1<r_0$.}
We have found the viscous torques in the disk that follow from the specified model of magnetic torques (see \S5). These, in turn, imply a luminosity profile of the disk, $dL_{\rm visc}/dr$.
{While we have found the dependence of $r_0$ on the magnetic dipole for a specific model of magnetic field interaction with the disk, it is remarkable that most of the disk quantities presented in this paper need have no explicit dependence on the magnetic field, and depend directly only on $\dot M$, $r_0$, and $r_c$. When expressed in a dimensionless form, 
{i.e., in combinations such as $r/r_c$ or $L/(GM\dot M/r_c)$,
}
the disk quantities depend on only one parameter, the ratio $r_0/r_c$, whose 3/2 power is the ``fastness" $\omega$.
}

While most authors agree (for a fixed value of the magnetic dipole) as to the value of the mass accretion rate, $\dot M_c\sim \dot M_0=\mu^2 r_c^{-7/2}(GM)^{-1/2}$, that corresponds to a termination radius of the disk that is more or less equal to the corotation radius, there is considerable theoretical disagreement as to what happens at lower accretion rates. 
A key question on which the authorities disagree is whether or not a strong magnetic field can prevent long-term accretion when the rate of flow of matter through the outer disk decreases below a critical value.

It is possible that at very low mass-accretion rates, say
$\dot M << \dot M_0/10$, the disk is expelled beyond corotation, and angular momentum is removed by outflowing matter (see the discussion in \S1).
Indeed, among  the intermediate polars (in which there is clear evidence for the presence of both strong magnetic fields and  an accretion disk) one system, AE Aqr, displays characteristics that may correspond to the propeller effect (Mikolajewski et al. 1996). Of course, for extremely strong stellar magnetic fields the disk may be disrupted completely, as in AM Her stars.

However, our results indicate that regardless of whether the disk is diamagnetic or not, and regardless of the detailed model of the disk--magnetosphere interaction,
for mass accretion rates that are somewhat lower than the critical value, $\dot M<\dot M_c$, there is a solution in which the disk terminates just within the corotation radius, $r_c$. This is because the stellar torques must remove angular momentum from the disk to allow its termination, while outside $r_c$ the magnetosphere can only deposit angular momentum in the disk (Wang 1987). 

At these, and comparable, accretion rates ($\dot M\sim \dot M_c$), while removing angular momentum from the disk just inside the corotation radius, the magnetic torques may deposit angular momentum at $r>r_c$, possibly at a substantial rate (depending on the torque model). This extra angular momentum must be transported outwards by viscous processes operating in the disk, leading necessarily to energy dissipation which increases the luminosity of the disk.
Thus, in this spin-down regime, we report a qualitatively new result, which seems to have been overlooked in the literature. The angular momentum deposited in the disk by the pulsar torques leads to a substantial additional dissipation of energy in the disk (beyond the value expected from gravitational release alone) as angular momentum is transported outwards by viscous processes. This leads to a change in the total luminosity of the disk, $L_{\rm vis}$, as well as of its radial distribution (eqs.~[\ref{elvis}],[\ref{elvint}]), possibly quite dramatic at lower accretion rates. Consequently, previous estimates of the mass accretion rate of spinning-down pulsars may be wrong by up to a factor of several (Fig.~7),
depending on the range of applicability of our model.

The total contribution of magnetic origin to luminosity, already considered by Kenyon et al. (1996) in their study of T-Tauri stars is the magnetic torque power, $L_{\rm tot} - GM\dot M/(2r_0)$, which includes the power $L_{\rm tot} - L_{\rm vis}$ released because of a mismatch in the rotation rate of the star and the parts of the accretion disk entrained by the magnetosphere (eqs.~[\ref{Ltota}], [\ref{Lmismatch}]). It is not clear {\it a priori} what fraction of the excess  $L_{\rm tot} - L_{\rm vis}$ is released inside the disk, and what fraction is released in non-thermal processes above the disk.

We have quantified these processes within the context of two alternate specific models of distributed magnetic torques, both of which correspond to no shielding and no diamagnetism. 
{Observations of torques as a function of luminosity may help in selecting the correct model of the magnetic field interaction with disk. For example, if the pulsed luminosity in X-ray pulsars is a good measure of the mass accretion rate, the observations discussed in \S \ref{sec:history} imply that in the spin-up phase the torque is proportional to $\dot M^\beta$, with $\beta>6/7$. In our model I, at high accretion rates the torque is
proportional to $\dot M^{9/10}$, while in the alternate prescription II (Appendix B) it is proportional to  $\dot M^{6/7}$.
However, the spin-up line for pulsars, eq.~(\ref{spinup}), seems not to be very sensitive to the detailed model of magnetic interaction.
}
The model of disk-magnetosphere interaction adopted in this paper does not 
quantitatively explain the reported rapid sign reversals of the torques in X-ray pulsars, with no apparent change in torque magnitude. However, already in our simple model the transition from spin-up to spin-down occurs over a fairly small luminosity range, and the spin-down torque quickly reaches an asymptotic value (Fig.~8), as the disk luminosity drops. This behavior has some qualitative resemblance to
observations. It may be worthwhile to repeat the calculations for a partially diamagnetic disk. 

The results reported here include an analytic solution for the radial profile of angular velocity, both in the region where the viscous stresses do not vanish
(which coincides with the region where the disk is Keplerian---this is the adopted ansatz), and inside the region where the viscous torques vanish.
The main conclusions of the paper, such as that regarding the enhanced luminosity of the disk for spinning-down pulsars, are valid for any thin disk model (e.g., for any viscosity prescription).  For a specific viscosity prescription, the $\alpha$-disk, we have also computed the complete analytic radial profiles of the disk variables $T$, $\rho$, $P$, and $H$ (\S7).

\bigskip\par\noindent
\centerline{\it Note added in manuscript}

        Since this paper was written the most extensive timing
analysis on an accretion-powered millisecond X-ray pulsar (SAX
J1808.4-3658) has been reported by Hartman et al.~(2007), who detect a
long-term (i.e., over 7 years) spin down of the X-ray pulsar of $\dot
\nu \simeq -5.6 \times 10^{-16}$ Hz s$^{-1}$.  Perhaps of greater
relevance to this paper, is the upper limit they set of $|\dot \nu|
\lesssim 2.5 \times 10^{-14}$ Hz s$^{-1}$, {\em during} a typical
X-ray outburst.  We have used the X-ray outburst profile of 2002 from
Fig.~3 of Hartman et al.~(2007) to numerically integrate our eq.~(36)
for the torque on the neutron star in order to determine the allowed
range of neutron star magnetic moment, $\mu$, that is consistent with
the upper limit on the spin changes during a typical outburst.  We
find that $ \mu \lesssim 3 \times 10^{26}$ G cm$^3$, 
corresponding to a surface magnetic field strength of 
$B \lesssim 3 \times 10^8$ G for a neutron star
with a radius of 10 km.  A similar value has been reported
by Burderi et al. (2006).
These are within the range of values cited by
Hartman et al.~(2007) for their constraint set during the outbursts;
however, our model for the torques is substantially different than
theirs.

\acknowledgements 

We thank Al Levine, Ed Morgan, Ron Remillard, Jon Arons and Marek Abramowicz for helpful discussions. WK would like to acknowledge the hospitality of colleagues at the Kavli Institute at MIT. { We are grateful to the anonymous referee for drawing our attention to the work of Campbell and Matthews et al., as well as for other helpful comments.} Antonia Savcheva's M.I.T. undergraduate senior thesis involved a number of studies related to this work.
SR acknowledges support from NASA Chandra Grant TM5-6003X.

\appendix
\section{Appendix A. Previous estimates of the magnetospheric radius}

We briefly review a number of the earlier estimates of the disk termination
radius that are found in the literature.
Let us begin by asking at what value of a frozen-in poloidal magnetic field  it would be energetically more favorable for plasma inside a flux tube to move out of the plane of the disk, rather than radially inwards.
This would occur whenever the magnetic field is larger than the one given by the expression
\bea
\label{Rees}
{B^2 \over 8\pi} = P,
\eea
where $P$ is the plasma pressure in the inner disk. This is the same condition that was derived by Pringle \& Rees (1972): ``the radius down to which the gas can crush the stellar field (...) is the radius at which the effective pressure in the disk is equal to the external magnetic pressure" [assuming that this radius is within the corotation radius].
Neglecting screening of the field one substitutes the magnetic dipole value
$B=\mu/r^3$ to obtain an estimate of this radius, $r_B$ through
\bea
{\mu^2 \over 8 \pi r_B^6} = P.
\eea

To compare other workers' results, we recall that in a standard thin 
accretion disk all velocities scale as the Keplerian velocity
$\vk= \sqrt{GM/r}$, multiplied by a power of the dimensionless disk thickness
$H/r<<1$ (Shakura \& Sunyaev 1973, Regev 1983, Klu\'zniak \& Kita 2000).
Thus, for example, the speed of sound scales as 
\bea
(P/\rho)^{1/2} \sim c_s \sim (H/r) \,\vk,
\eea
while the radial drift velocity is
\bea
v_r \sim \alpha (H/r)^2 \vk.
\eea
Here, $\rho$ is the fluid density, and $\alpha$ is a dimensionless viscosity parameter. The radial angular momentum flux is then 
\bea
\rho v_r \vk  r \sim \alpha \rho c_s^2 r \sim \alpha P r.
\eea

In the Ichimaru (1978) model, the disk is Keplerian everywhere except in a narrow boundary layer, in which all excess angular momentum is removed, so that at the base of the boundary layer, at $r=r_A$, the fluid can come to rest in the stellar frame. The condition for hydrostatic equilibrium
\bea
{B^2 \over 8\pi} = {GM \sigma \over r^2_A},
\eea
with $\sigma \sim \rho l$ the surface density in the boundary layer of
width $l$,
translates into
\bea
 {B^2 \over 8\pi}\sim (l/l_{\rm BL})\,  P,
\eea
where $l_{\rm BL} \sim (H/r)^2 r_A$ would be the thickness of a standard hydrodynamic boundary layer at this radius  (Pringle 1977).

Scharlemann (1978) considers a completely diamagnetic disk and estimates the condition for stress balance between the disk and the screening toroidal current as
\bea
B^2 \sim 21\, P.
\eea
We note that this is within the range of the other estimates discussed
here, as $4\pi < 21 < 8 \pi$.

Arons (1993) considers a diamagnetic disk penetrated by the (enhanced) dipole over a region of radial extent comparable to the disk thickness, and writes the stress balance at the disk termination radius $r=r_m$ as
\bea
 {B_z B_\phi \over 4\pi} \approx \rho v_r v_\phi.
\eea
With $v_\phi \approx \vk$, and $B_z \sim B_\phi$ this yields 
\bea
 {B^2 \over 4\pi} \approx \alpha P.
\eea
However, following the exact solution of Aly (1980) for the field configuration outside a diamagnetic disk, Arons takes the strength of the magnetic field to be enhanced by a factor of $(r/H)^{1/2}$ over the dipole value,
yielding
\bea
\label{Arons}
 {\mu^2 \over 4\pi r_m^6} \approx \alpha (H/r_m) P.
\eea
Actually, Arons (1993) eliminates $\rho v_r$ with the help of the mass conservation equation for a steady disk, $\dot M \approx 4 \pi H r \rho v_r$,
to obtain a final condition
on the inner radius of the disk in terms of the stellar mass accretion rate $\dot M$:
\bea
\label{Aronsr}
r_m = \left( GM\right)^{-1/7}\dot M^{-2/7} \mu^{4/7} 
~~ .
\label{arons}
\eea
We have adopted this definition of $r_m$ as a convenient fiducial length.

Wang (1987) obtained the same value (up to a factor of $2^{1/7}$) by considering the condition that the magnetic stresses be able to remove enough angular momentum of the disk:
\bea
B_\phi B_z r^2 = \dot M \frac{d}{dr} (\vk r)
 = {1 \over 2} \dot M \vk.
\label{Wang}
\eea
Wang (1987) expressly states that eq.~(\ref{Wang}) holds only if the
resulting  termination radius is less than the corotation radius---otherwise,
the central dipole would be unable to remove angular momentum from the disk.
Again, with $B_\phi \sim B_z$, where $B_z$ is the unscreened and un-enhanced dipole value of the magnetic field, this would yield $r=2^{1/7}r_m$ as the disk termination radius.
However, Wang (1987, 1995) uses a more complicated prescription for the value of $B_\phi$, leading to an estimate very similar to our eq.~(\ref{bc}) in the text. 

Clearly, as $\alpha \cdot (H/r)<<1$, the condition of eqs.~(\ref{Arons}) and
 (\ref{arons}) holds before the Pringle-Rees value of the magnetic field,
i.e., eq.~(\ref{Rees}), can be attained. However, $r_B$ is smaller than $r_m$ only by a factor 
$\sim(\alpha H/r)^{1/7} \approx 0.4$, where for the numerical estimate we took 
$\alpha \approx 0.01$ and $H/r \approx 0.1$.

In a very influential paper, Ghosh \& Lamb (1979) suggested that a large region of the disk is threaded by a screened magnetic field of $\sim0.2$ the strength of the external dipole, or less, and claim that the disk
terminates at the radius
\bea
r_{\rm GL} \approx 0.41\cdot 2^{-1/7} r_m.
\eea
One concern may be that the value $r_{\rm GL} = 0.37 r_m$ is lower than the Pringle-Rees limit $r_B$ derived from eq.~(\ref{Rees}), and indeed,  Ghosh \& Lamb assume that matter leaves the disk vertically at a rate specified by an arbitrary ``gating function."
A number of authors (Pringle \& Rees 1972; Ichimaru 1978; Wang 1987; Spruit \& Taam 1993; Rappaport, Fregau \& Spruit 2004) restrict the termination radius to be within the corotation radius. 

In summary, the various authors differ by no more than a factor of 2 or 3
in their estimate of radius at which the central magnetosphere terminates
a Keplerian accretion disk,
when this termination radius is within the corotation radius.
Before the magnetic pressure becomes sufficiently strong to channel matter away
from the symmetry plane of the disk, magnetic torques acting on the inner
disk attain a value sufficient to remove angular momentum of the accreting
matter at the rate corresponding to its advection
at the inner edge of the disk,
$\dot M v_{\rm K} r$.

\section{Appendix B. Results for alternate prescription for $B_\phi$}
\label{AppB}
Most of the results given in the text have been for the prescription where $B_\phi$ is given by
\bea
B_\phi^I \simeq  B_z \left( 1 - \frac{\Omega}{\omega_s} \right) ~~,
\eea
for all radii, i.e., $r<r_c$ and $r\ge r_c$. 
Here we give the principal analytic results for the alternate prescription for $B_\phi^{II}$ that we have considered:
{
\bea
\label{prescrII}
B_\phi^{II} &\simeq & B_\phi^I (\omega_s/\Omega)   \simeq -B_z \left( 1 - \frac{\omega_s}{\Omega} \right)~~{\rm for}~~r<r_c ~~,\\
B_\phi^{II} &\simeq & B_\phi^I ~~{\rm for}~~r\ge r_c ~~
\eea
}
and for $r\le r_c$ the last term in eq.~(\ref{visc1}) is then replaced by an alternate form of the magnetic torque
\begin{equation}
\label{torqueII}
- \frac{B_z^2 r}{4\pi H} \left( 1 - \frac{\omega_s}{\Omega} \right)~~.
\end{equation}
For the region $r>r_c$ the prescription remains the same as for $B_\phi^{I}$.
The corresponding vertically integrated $r$ - $\phi$ component of the viscous stress-energy tensor, analogous to eq.~(\ref{visc2}) in the text is:
{\small
\bea
\label{TrphiII}
-T_{r\phi} &=& \frac{\dot M\sqrt{GM}}{2\pi r^2} \left(\sqrt{r} -\sqrt{r_0} \right)
+ \frac{\mu^2}{6 \pi r^5}    \left[2\sqrt{\frac{r^3}{r_c^3}}\left(\sqrt{\frac{r^3}{r_0^3}} - 1\right) + 1-\left(\frac{r}{r_0}\right)^3 \right]~~~{\rm for}~~~ r \le r_c~,  \\
-T_{r\phi} &=&\frac{\dot M\sqrt{GM}}{2\pi r^2} \left(\sqrt{r} -\sqrt{r_0} \right) + \frac{\mu^2}{18 \pi r^5}    \left[-2\left(\frac{r}{r_c}\right)^3  +2 \left(\frac{r_c}{r}\right)^{3/2} +\frac{6r^3}{\left(r_0 r_c\right)^{3/2}} -3\left(\frac{r}{r_0}\right)^3 -3\right]~~~{\rm for}~~~ r \ge r_c~.
\eea
}
The boundary condition for the inner edge of the viscous accretion disk, analogous to eq.~(\ref{bc}), is (Wang 1995):
\bea
\label{bcbis}
1 = 2 \left(\frac{r_m}{r_0}\right)^{7/2} \!\! \left(1-\sqrt{\frac{r_0^3}{r_c^3}}\right) = 2 \xi^{7/2}  \omega^{-7/3} \left(1-\omega\right)~~, 
\eea
 The asymptotic forms of eq.~(\ref{bcbis}) are 
\bea
\label{bcII}
r_0/r_c \simeq  2^{2/7}\xi ~~~~{\rm for}~~\xi \ll 1~~,
\eea
\bea
r_0/r_c \simeq 1- \frac{1}{3} \xi^{-7/2} ~~~~{\rm for}~~\xi \gg 1
\eea
which is the same as eq.~(\ref{r0}). 
{For a system with fixed $\mu$, $\omega_s$, and $M$, eq.~(\ref{bcII}) implies that $r_0\propto r_m\propto\dot M^{-2/7}$ in the high mass accretion rate limit of eq.~(\ref{bcbis}).
}
The counterpart of eq.~(\ref{bcfast}) is
\bea
\label{bcIIfast}
\left(r_m/r_0\right)^{7/2}  = [2(1-\omega)]^{-1}~,
\eea
{or $2\dot M_0(1-\omega)= \dot M \omega^{7/3}$,
}
and the same limits obtain as for eq.~(\ref{bcfast}): for $\dot M\rightarrow 0$, $\omega \rightarrow 1$; for $\mu\rightarrow 0$, $\omega\rightarrow 0$. However, now $r_m/r_0\rightarrow 1/2^{2/7}$ (and not $0$) as $\mu\rightarrow 0$.
The expression for $dL_{\rm vis}/dr$ corresponding to eq.~(\ref{elvis}) in the text is given by:
{\footnotesize
\bea
\frac{dL_{\rm vis}}{dr} &=& \frac{3GM\dot M}{2r^2} \left\{ \left(1 -\sqrt{\frac{r_0}{r}} \right) + \frac{1}{3}  \left({\frac{r_m}{r}}\right)^{7/2} \left[2\sqrt{\frac{r^3}{r_c^3}}\left(\sqrt{\frac{r^3}{r_0^3}} - 1\right) + 1-\left(\frac{r}{r_0}\right)^3 \right]   \right\}~~{\rm for}~~ r \le r_c~,\\
\frac{dL_{\rm vis}}{dr} &=& \frac{3GM\dot M}{2r^2} \left\{ \left(1 -\sqrt{\frac{r_0}{r}} \right) + \frac{1}{9} \left({\frac{r_m}{r}}\right)^{7/2} \left[-2\left(\frac{r}{r_c}\right)^3  +2 \left(\frac{r_c}{r}\right)^{3/2} +6\left(\frac{r^2}{r_0 r_c}\right)^{3/2} -3\left(\frac{r}{r_0}\right)^3 -3\right]   \right\}\nonumber\\&&~~~{\rm for}~~~ r \ge r_c~.
\eea
}
The total torque on the central star, analogous to eq.~(\ref{taus}) in the text is given by:
\bea
 \tau_{\rm CS}= \dot M \left(\sqrt{GMr_0} - \omega_s R_s^2 \right) -\frac{\mu^2}{9r_0^3}  \left[-2\left(\frac{r_0}{r_c}\right)^3 +6\left(\frac{r_0}{r_c}\right)^{3/2}-3\right]~~ ,
 \eea
{For $R_s\rightarrow0$ and $\dot M\ne0$, this reduces to
\bea
\label{dimlesstorqueII}
 \tau_{\rm CS}= \dot M \sqrt{GMr_0}\, \left[7/6- (4/3)\omega+(1/9)\omega^2\right]/(1-\omega)
\eea
(Wang, 1995).
The dimensionless torque corresponding to eq.~(\ref{dimlesstorqueIalt}) is given by $g_{II}(\omega)$, with
\bea
\label{dimlesstorqueIIb}
g_{II}(\omega)=\frac{7}{6}\left[1- \frac{1}{21} \frac{3 \omega-2\omega^2}{1-\omega}\right]~~.\eea
The first root is at $\omega=[(12-\sqrt{102})/2]\approx 0.9502\approx (0.9666)^{3/2}$, quite close to the root of $g_I$ in prescription I (eq.~[\ref{dimlesstorque}]): 20/21.
}

\par\noindent
Finally, the factor $F$, corresponding to eq.~(\ref{F}) is
{\footnotesize
\bea
\label{FII}
F &=&  \left(1 -\sqrt{\frac{r_0}{r}} \right) + \frac{1}{3}  \left({\frac{r_m}{r}}\right)^{7/2} \left[2\sqrt{\frac{r^3}{r_c^3}}\left(\sqrt{\frac{r^3}{r_0^3}} - 1\right) + 1-\left(\frac{r}{r_0}\right)^3 \right] ~~{\rm for}~~ r \le r_c~, \\
\label{FIIb}
F &=&  \left(1 -\sqrt{\frac{r_0}{r}} \right) + \frac{1}{9}   \left({\frac{r_m}{r}}\right)^{7/2} \left[-2\left(\frac{r}{r_c}\right)^3+2 \left(\frac{r_c}{r}\right)^{3/2} +6\left(\frac{r^2}{r_0 r_c}\right)^{3/2} -3\left(\frac{r}{r_0}\right)^3 -3\right]~~{\rm for}~~r \ge r_c~.\nonumber\\&& ~~~
\eea
}
\par\noindent
{ To within a factor of two, the magnetic torque of eq.~(\ref{torqueII}) was adopted by Matthews et al. (2005) for all $r$, accordingly, to within the same factor of two, our eq.~(\ref{bcbis}) is the same as their eq.~(26),
although its interpretation is different (see our discussion of eq.~[\ref{bc}] in \S~\ref{sec:match}). After the substitution of parameter values $\beta \rightarrow \mu^2/(\pi\sqrt{GM})$, $\gamma=7/2$, eqs.~(30-33) of Matthews et al. (2005) coincide with our eq.~(\ref{FII}) in the region of overlap of the torque prescription in their paper and this Appendix ($r \le r_c$). 
However, for $r > r_c$, we obtain a different result, eq.~(\ref{FIIb}), because Matthews et al. have adopted a different $B_\phi$ prescription for this regime.

{After using eq.~(\ref{bcIIfast}) we obtain the function $F(\sqrt{r_0/r},\omega)$ that appears in the $\alpha$-disk equations
\bea
\label{FIIfast}
F(y,\omega) & = &[6(1-\omega)]^{-1}\left(6 -7y + y^7 -6\omega + 8\omega y - 2\omega y^{4}\right)~,~{\rm for}~~\omega^{1/3}\le y\le 1~,\\
\label{FIIbfast}
F(y,\omega) & = &[18(1-\omega)]^{-1}\left(18 - 21y - 3y^7 - 18\omega + 24\omega y -2\omega^2 y + 2\omega^{-1} y^{10}\right),\nonumber\\
&&~~~~{\rm for}~ 0<y\le \omega^{1/3}~.
\eea
Again $F(1,\omega)=0$, and the zero magnetic field limits of eqs.~(\ref{FIIfast}), (\ref{FIIbfast}) are $F(y,0)=1-y-\left(y-y^7\right)/6$, and $F(y,0)=1-y-\left(y+y^7\right)/6\rightarrow1$, respectively. Note that in view of the domain restriction $y\le \omega^{1/3}$, the last term in eq.~(\ref{FIIbfast}), i.e., $\omega^{-1} y^{10}$, is not divergent when $\omega\rightarrow0$.
}

\section{Appendix C. A magnetically dominated Keplerian accretion disk}

{
The specific prescription for $B_\phi$ that we are using, is incompatible with a magnetically dominated Keplerian disk, i.e., the solution to eq.~(\ref{novisc}) is not Keplerian. This prescription is equivalent to assuming the following  form for magnetic diffusivity: $\eta_1\sim rH\Omega$. Campbell (1992) uses a different form, $\eta_{\rm C}=v_{\rm A}H$, where the Alfv\'en speed $v_{\rm A}\sim B_\phi/(4\pi\rho)^{1/2}$. His work cannot be directly compared with ours, because we restrict ourselves to a discussion of a thin accretion disk. Eq.~(53) of hydrostatic equilibrium in Campbell (1992) neglects gravity, but for a gravitationally stratified disk with sound speed, $\sim(P/ \rho)^{1/2}\simeq \Omega_KH$, equal to the Alfv\'en speed $v_{\rm A}$, Campbell's diffusivity would differ from ours by a factor of $H/r$:
$\eta_{\rm C}\simeq\eta_1(H/ r)$. [In fact, in the Campbell (1992) solution, $H\sim r$, so that in terms of their numerical values $\eta_{\rm C}\sim\eta_1$.]
}

{ Inspired by Campbell's approach, in this Appendix we discuss the solution for a gravitationally stratified, Keplerian accretion disk that is dominated by magnetic torques.
We take $B_z=\mu r^{-3}$, $\dot M = 4\pi r v_r\rho H$, $\eta_C=\Omega_KH^2$,
\bea
\label{Alfven}
 B_\phi/(4\pi\rho)^{1/2} = v_{\rm A} =\Omega_KH,
\eea
\bea
\label{haha}
-B_\phi=rHB_z(\Omega_K - \omega_s)/\eta_C = B_z(r/H)(1- \omega_s/\Omega_K),
\eea
and consider the region within the corotation radius, $\Omega_K > \omega_s$.

The angular momentum conservation equation
\bea
\label{Ctorque}
-\dot M \frac{d}{dr}\left(r^2\Omega_K\right) = \frac{\mu}{r} B_\phi
\eea
implies that
\bea
B_\phi =\frac{\dot M\Omega_K r^2}{2\mu} \propto r^{1/2}
\eea
The height of the disk now follows directly from eq.~(\ref{haha}),
while the density from eq.~(\ref{Alfven})
\bea
H &=& \frac{\mu^2}{\dot M\Omega_K r^4}\left(1- \frac{\omega_s}{\Omega_K}\right) \propto r^{-5/2}\left(1- \frac{\omega_s}{\Omega_K}\right),\\
\rho &=& \frac{\dot M^4\Omega_K^2r^{12}}{4\pi\mu^6} \left(1- \frac{\omega_s}{\Omega_K}\right)^{-2} \propto r^9 \left(1- \frac{\omega_s}{\Omega_K}\right)^{-2}.
\eea
Note the divergence in density at the corotation radius.

The radial velocity starts from zero at the corotation radius, and rapidly increases inwards:
\bea
\label{velocity}
v_r \propto (rH\rho)^{-1} \propto r^{-15/2}\left(1- \frac{\omega_s}{\Omega_K}\right).
\eea
Since all of the angular momentum is removed locally from the disk by the magnetic torques, the torque on the star from an annulus at $r$ is $(1/2)\Omega_K\dot Mr\, dr$, see eq.~(\ref{Ctorque}), and the power applied by the annulus to spin up the star is $(1/2)\dot M\Omega_K\omega_sr dr$.
At the corotation radius all the power of the disk is used to torque the star, and so the disk is dark; for smaller radii the disk becomes brighter as the relative amount of angular momentum removed decreases. The radiated flux is then 
\bea
\label{fluxC}
2F_\gamma=\frac{\dot M}{4\pi}\Omega_K^2\left(1- \frac{\omega_s}{\Omega_K}\right),
\eea
the first term being the rate of energy change of the fluid over unit area
\bea
\frac{\dot M}{2\pi r} \frac{d}{dr}\left(-\frac{1}{2}\frac{GM}{r}\right) =\frac{\dot M}{4\pi}\Omega_K^2,
\eea
the second term is the spin-up power drained from the disk per unit area.
This expression, eq.~(\ref{fluxC}), agrees with the Joule heating formula derived by Campbell (1992):
\bea
2F_\gamma=2r^2B_z^2(\Omega_K - \omega_s)^2H/\eta_C.
\eea
The total torque on the star is 
\bea
\frac{1}{2}\int_0^{r_c}\dot M\Omega_Krdr =\dot M l_K(r_c),
\eea
assuming $R_S \ll r_c$. Here, $l_K\equiv\Omega_K r^2$ is the Keplerian specific angular momentum.

We conclude that with a suitable prescription for a magnetic field, it is possible to construct a magnetically dominated accretion disk that {\sl formally} has Keplerian orbital velocities. In fact, however, the trajectories of the fluid would not be close to Keplerian orbits for all radii---in addition to its azimuthal motion, the fluid suffers rapid radial inflow. Because the radial velocity is controlled directly by the prescribed magnetic field---at steady accretion rate, eq.~(\ref{velocity}) ultimately follows from eq.~(\ref{haha}) and from the equation of vertical hydrostatic equilibrium (eq.~[\ref{Alfven}])---the solution  cannot be considered valid for all $r<r_c$ until its consistency with the radial equation of motion has been checked. 
}

{
In the main body of this paper, we find a solution for the angular velocity profile in the magnetically dominated region which is not Keplerian at all (Fig.~\ref{fig:matching}). The counterpart of eq.~(\ref{fluxC}) for the region $r<r_0$ is
\bea
\label{fluxB}
{\cal F} = \frac{\dot M}{2\pi r} \left[\left(\Omega_K^2 - \Omega^2\right)r + \frac{d}{dr} \left(\Omega r^2\right)(\Omega - \omega_s)\right].
\eea
This is the power released per unit area of the disk over and above the mechanical power used to torque up the star. The first term in parentheses in eq.~(\ref{fluxB}) is the power used to accelerate the fluid radially. For $\Omega=\Omega_K$, eq.~(\ref{fluxB}) reduces to eq.~(\ref{fluxC}).
}
}

\end{document}